\documentclass[iop]{emulateapj}
\usepackage{graphicx}
\usepackage{aas_macros}
\usepackage {amsmath}
\usepackage{amssymb}
\usepackage{graphics}
\usepackage{wrapfig}
\usepackage{soul}
\usepackage{natbib}
\usepackage{array}
\usepackage[export]{adjustbox}
\usepackage{multirow}
\usepackage[table]{xcolor}
\usepackage{booktabs,colortbl,tabularx}
\newcolumntype{P}[1]{>{\centering\arraybackslash}p{#1}}
\newcolumntype{M}[1]{>{\centering\arraybackslash}m{#1}}
\usepackage{comment}
 \slugcomment{Accepted to \apj}

\newcommand{\um}{\ensuremath{{\mu}}m\;}
\newcommand{\mum}{\ensuremath{{\mu}}m}

\newcommand{\ox}{[{\ion{O}{3}}]}
\newcommand{\oii}{[{\ion{O}{2}}]}
\newcommand{\bluox}{[\ion{O}{3}]~$\lambda$4960}
\newcommand{\redox}{[\ion{O}{3}]~$\lambda$5008}
\newcommand{\nii}{[{\ion{N}{2}}]}
\newcommand{\sii}{[{\ion{S}{2}}]}

\newcommand{\blun}{[\ion{N}{2}]~$\lambda$6550}
\newcommand{\redn}{[\ion{N}{2}]~$\lambda$6585}
\newcommand{\hb}{{H$\beta$}}
\newcommand{\ha}{{H$\alpha$}}
\newcommand{\ybpt}{[{\ion{O}{3}}]/{{H$\beta$}}}
\newcommand{\xbpt}{[{\ion{N}{2}}]/{{H$\alpha$}}}

\newcommand{\sfratio}{$(\rm SFR_{\rm IR}+SFR_{\rm UV})/\rm SFR_{ \rm UV,corr}$ \;} 

\newcommand{\irxg}{MIR-excess galaxies\;} 
\newcommand{\irx}{MIR-excess\;} 
\newcommand{\lir}{$\rm L_{\rm IR}$\;}
\newcommand{\lumir}{$\rm L_{\rm IR}$}
\newcommand{\lsun}{$\rm L_{{\odot} }$\;}

\newcommand{\z}{$z\sim2$\;}
\newcommand{\zz}{$z\sim2$}
\newcommand{\da}{\cite{Daddi2007a}\;}
\newcommand{\db}{\cite{Daddi2007b}\;}

\newcommand{\rang}{\cite{Rangel2013}\;}
\newcommand{\sfir}{$\rm SFR_{\rm IR}\;$}
\newcommand{\sfuv}{$\rm SFR_{UV}$\;}
\newcommand{\sfuvcorr}{$\rm SFR_{\rm UV,corr}$\;}
\newcommand{\leight}{$\rm L_{ \rm 8}\;$}
\newcommand{\reight}{$\rm <$$ \rm L_{IR}$$>$/$\rm <$$\rm L_{8}$$>$\;}
\newcommand{\sfha}{$\rm SFR_{\rm H\alpha}\;$}

\newcommand{\e}[1]{{#1}}
\newcommand{\rf}[1]{{#1}}
\newcommand{\ma}[1]{\textcolor{black}{\e{#1}}}

\shorttitle{The Nature of Mid-Infrared Excess Galaxies}
\shortauthors{Azadi et al.}

\begin{document}

\title{THE MOSDEF SURVEY: The Nature of Mid-Infrared Excess Galaxies and a Comparison of IR and UV Star Formation Tracers at \z}

\author{
Mojegan Azadi \altaffilmark{1,2}, 
Alison Coil \altaffilmark{2},
James Aird \altaffilmark{3},
Irene Shivaei \altaffilmark{4},
Naveen Reddy \altaffilmark{5},
Alice Shapley \altaffilmark{6},
Mariska Kriek \altaffilmark{7},
William R. Freeman \altaffilmark{5},
Gene C. K. Leung \altaffilmark{2}, 
Bahram Mobasher \altaffilmark{5},
Sedona H. Price \altaffilmark{8},
Ryan L. Sanders \altaffilmark{6},
Brian Siana  \altaffilmark{5}, and
Tom Zick \altaffilmark{7}
}
\altaffiltext{1}{Harvard-Smithsonian Center for Astrophysics, 60 Garden Street, Cambridge, MA, 02138, USA}
\altaffiltext{2}{Center for Astrophysics and Space Sciences, Department of Physics, University of California, 9500 Gilman Dr., La Jolla, San Diego, CA 92093, USA}
\altaffiltext{3}{Department of Physics \& Astronomy, University of Leicester, University Road, Leicester LE1 7RH, UK}
\altaffiltext{4}{Steward Observatory, University of Arizona, Tucson, AZ, 85721, USA}
\altaffiltext{5}{Department of Physics and Astronomy, University of California, Riverside, 900 University Avenue, Riverside, CA, 92521, USA}
\altaffiltext{6}{Department of Physics and Astronomy, University of California, Los Angeles, 430 Portola Plaza, Los Angeles, CA 90095, USA}
\altaffiltext{7}{Astronomy Department, University of California at Berkeley, Berkeley, CA 94720, USA}
\altaffiltext{8}{Max-Planck-Institut für extraterrestrische Physik, Postfach 1312, Garching, 85741, Germany}

\begin{abstract} 

We present an analysis using the MOSFIRE Deep Evolution Field (MOSDEF) survey on the nature of ``MIR-excess" galaxies, which have star formation rates (SFR) inferred from mid-infrared (MIR) data that is substantially elevated relative to that estimated from dust-corrected UV data. We use a sample of $\sim$200 galaxies and AGN at $1.40<z<2.61$ with 24 \um detections \rf{(rest-frame 8\mum)} from MIPS/\textit{Spitzer}. We find that the identification of \irx galaxies strongly depends on the methodologies used to estimate IR luminosity (\lumir) and to correct the UV light for dust attenuation. We find that extrapolations of the SFR from the observed 24 \um flux, using luminosity-dependent templates based on local galaxies, substantially overestimate \lir in $z\sim2$ galaxies. By including \textit{Herschel} 
observations and using a stellar mass-dependent, luminosity-independent \lir, we obtain more reliable estimates of the SFR and a lower fraction of \irx galaxies. Once stellar mass selection biases are taken into account, we identify $\sim24\%$ of our galaxies as MIR-excess. However, \sfha is not elevated in \irx galaxies compared to MIR-normal galaxies, indicating that the intrinsic fraction of \irx may be lower. Using  X-ray, IR, and optically-selected AGN in  MOSDEF, we do not find a higher prevalence for AGN in MIR-excess galaxies relative to MIR-normal galaxies. A stacking analysis of X-ray undetected galaxies does not reveal a harder spectrum in MIR-excess galaxies relative to MIR-normal galaxies. Our analysis indicates that AGN activity does not contribute substantially to the MIR excess and instead implies that it is likely due to the enhanced PAH emission.

\end{abstract}

\keywords{ galaxies: Mid-IR excess -- galaxies: high-redshift -- infrared: galaxies -- galaxies: active}

\section{Introduction}
\label{sec:intro} 

In the study of galaxies, the  star formation rate (SFR) is a fundamental quantity that provides valuable information about the underlying physical processes that drive galaxy  evolution. As the energy output from stars is radiated over a wide range of wavelengths from X-ray to radio, the SFR of galaxies can be traced at various wavelengths \citep[see e.g.][]{kennicutt1998star,Kennicutt2009,Hao2011,Calzetti2013}.

Ultraviolet (UV) radiation directly traces the population of young, massive stars in a galaxy and can thus provide a robust tracer of the SFR \citep[e.g.][]{kennicutt1998star,Calzettirev,Reddy2010,Magdis2010}.
However, to reliably estimate the total SFR from the observed UV radiation one must account for the effects of dust extinction \citep[e.g.][]{Meurer1999,Bouwens2009,Reddy2012,reddy2015}.

\ha\; line emission, originating from ionized nebular gas that re-emits the incident stellar luminosity, is another excellent SFR tracer. While less direct than the UV, \ha\ is less affected by dust extinction and thus can provide an accurate and reliable SFR tracer \citep[e.g.][]{Glazebrook1999,Kewley2002,Reddy2010,Shivaei2015}.

Infrared (IR) radiation traces the thermal emission of dust in a galaxy, which is primarily heated by young stars. 
The IR emission thus traces the SFR for the dust-obscured stellar populations within a galaxy. 
With the advent of IR satellites such as the Infrared Astronomical Satellite \citep[IRAS;][]{Neugebauer1984}, the Infrared Space Observatory \citep[ISO;][]{Kessler1996} and later \textit{Spitzer} \citep{Werner2004}, the \textit{Wide-field Infrared Survey Explorer} \citep[WISE;][]{Wright2010} and \textit{Herschel} \citep{Pilbratt2010}, remarkable progress has been made in our understanding of the stellar content and SFRs of galaxies. 

The Multi-band Imaging Photometer on \textit{Spitzer} \citep[MIPS;][]{Rieke2004} provides sensitive imaging at Mid-IR (MIR) wavelengths.The high sensitivity of the MIPS 24 \um band in particular allows us to directly detect moderately dusty star-forming galaxies out to high redshifts ($z\gtrsim2$), which may not be detectable at longer wavelengths \citep[e.g.][]{LeFloch2009,Lee2010}.


Many studies have investigated the agreement between IR, UV, and \ha\ SFR tracers \citep[e.g.][among many others]{Gonzalez2002,Kewley2002,Daddi2007a, Magdis2010,Reddy2010,Wuyts2011,Shivaei15b}. 
A number of studies use MIPS 24~\um data in particular to estimate the total IR luminosity of \z\ galaxies, providing SFR estimates that correlate well with other star formation tracers such as \ha\ \citep[e.g.][]{reddy2006,Shipley2016}.

However, several studies have found that there are \z galaxies for which the observed 24 \um flux is substantially elevated relative to what is expected from multi-wavelength data \citep[e.g.][]{Daddi2007a,Papovich2007}, which has resulted in the identification of so called ``\irx" galaxies. Specifically, \cite{Daddi2007a} report the discovery of a large population ($\sim$25\%) of \z BzK-selected galaxies with a \irx such that their combined IR and UV SFR exceeds the extinction-corrected UV SFR by a factor of three. However, the fraction of \irx galaxies identified by \db may depend on the accuracy of both the IR and UV star formation estimates, and there are known uncertainties on each that need to be treated carefully.


It has been shown that the identification of \irx galaxies depends on the template set that is used to estimate the total IR luminosity, \lumir. 
Several studies provide empirically- (or semi-empirically-) derived IR templates based on local star-forming galaxies \cite[e.g.][]{CE01,DH02,Rieke2009}.
\da used the luminosity-dependent templates from \cite{CE01} to estimate \lir based on the 24 \um data for \z galaxies. 
\cite{Salim2009}, however, used the \cite{DH02} templates to estimate \lir for galaxies at $z<1.4$ and identify a much lower fraction of \irx galaxies than \cite{Daddi2007a,Daddi2007b}. 
After the launch of \textit{Herschel}, which enabled the detection of IR luminous high redshift galaxies directly at far-IR (FIR) wavelengths, studies found that
the luminosity-dependent templates created based on local galaxies systematically overestimate \lir in galaxies with  $L_{IR}>10^{12} \; {L}_\odot$ at $z>1.5$ \citep[e.g.][]{Nordon2010,Muzzin2010,Elbaz2011}. 
This is due to high-redshift luminous galaxies having different dust temperatures than their local counterparts \citep[e.g.][]{Symeonidis2011}. To resolve the problem of overestimating \lir, more recent studies propose a single luminosity-independent conversion for \lir estimation from MIR observations \citep{Wuyts2008,Wuyts2011,Elbaz2011,Nordon2012,Reddy2012}. Furthermore, several studies indicate that stellar mass, metallicity, or specific star formation rate (sSFR) can also impact the \lir  estimation from 24 \um observations \citep[e.g.][]{Smith2007,Nordon2012,Utomo2014,Shivaei2017,Schreiber2018}.


The identification of \irxg also depends on the method used to correct the unobscured UV SFR for dust. \sfuv can be corrected for dust absorption using measurements of the UV continuum slope or using an average correction based on different attenuation curves \citep[e.g.][]{Cardelli1989,calzetti2000,reddy2015}. 
It has recently become possible using NIR spectroscopic data to directly measure the ratio of Balmer emission lines \ha/\hb\ (the ``Balmer decrement") in statistical samples of individual galaxies at $z\sim2$ \citep[e.g.][]{price2014,reddy2015,Nelson2016}. The Balmer decrement is an indicator of reddening of the ionized gas, which can also be used to estimate the reddening of the stellar continuum \citep[e.g.][]{reddy2015}. However, in dusty galaxies reddening corrections are very uncertain \citep[e.g.][]{Goldader2002}, and $L_{IR}/L_{UV}$ varies with stellar mass, dust temperature, and evolution of the dust temperature with redshift \citep[e.g.][]{Bouwens2016}. Underestimating the reddening correction will lead to underestimates of the dust-corrected SFR from the UV and thus to the misclassification of galaxies as MIR-excess galaxies.


Some studies have suggested that the elevated MIR flux in \irxg may be due to a contribution from a hidden population of active galactic nuclei (AGN) that are not identified at other wavelengths \citep[e.g.][]{Daddi2007b,Alexander2011,Rangel2013}. \db find that the spectral energy distribution (SED) of the \irxg peaks at redder wavelengths than the rest-frame 1.6 \um stellar peak. They also find a redder K-5.8 \um color in \irxg than in normal galaxies, which they conclude is due to an underlying AGN contribution. By performing an X-ray stacking analysis on MIR-excess galaxies without an X-ray detection, \cite{Daddi2007b} find a much harder X-ray spectrum in \irxg than in normal galaxies, which suggests the presence of Compton-thick AGN \citep[see also][]{Fiore2008}. 

More recently, \cite{Rangel2013}
used deeper \textit{Chandra} observations of BzK-selected galaxies in the  \textit{Chandra} Deep Field North (CDFN) and \textit{Chandra} Deep Field South (CDFS) fields and found no significant evidence for an elevated X-ray detection rate in \irxg compared to normal galaxies. Stacking analysis in \cite{Rangel2013} show a softer signal than \db which they argue is due to a mixture of highly absorbed AGN, low-luminosity unobscured AGN and star-forming galaxies. They conclude that the hard spectrum in the stacking analyses of earlier studies \citep[e.g.][]{Daddi2007b,Fiore2008} is due to a few hard X-ray sources that were just below the detection limit of the \textit{Chandra} data available at the time, which are now directly detected with deeper X-ray observations. \cite{Rangel2013} also find that while there are AGN among the \irx galaxy population, they are not typically Compton-thick AGN. They conclude that the bulk of the \irx galaxy population are luminous, dusty, star-forming galaxies where the SFRs are either overestimated in the MIR, underestimated in the UV, or both.

The depth and sensitivity of X-ray data clearly plays an important role in the rate of AGN identification \citep[e.g.][]{mendez2013primus,Azadi2017} and thus can affect conclusions regarding the nature of \irx\ galaxies.
AGN can also be identified at other wavelengths, such as a power-law like emission in the MIR \citep[e.g.][]{Lacy2004,Stern2005,donley2012,mendez2013primus} or optical diagnostics such as the ``BPT diagram" \citep[\ybpt\ versus \xbpt,][]{baldwin1981}. 
Using  multi-wavelength AGN identification techniques can therefore provide a clearer picture of the AGN contribution in \irx galaxies.  

\rf{We note that the excess 24 \um emission in \z galaxies could be due to enhanced emission from polycyclic aromatic hydrocarbon (PAH) molecules rather than obscured AGN activity \citep [e.g.][]{lutz2011pacs,Nordon2012}. Emission from PAH molecules is significant at rest-frame wavelengths of 5--12 \mum, but the strength of the emission varies with stellar mass and metallicity \citep[e.g.][]{Engelbracht2006, Smith2007, Shivaei2017} and may contribute up to 20\% of the \lumir\ \citep[e.g.][]{Smith2007}. Therefore, investigations of \irx galaxies should take into account the possible enhancement at \irx wavelengths due to PAH features.}


In this paper we study the nature of \irx galaxies at \z in the MOSFIRE Deep Evolution Field (MOSDEF) survey \citep{Kriek2015}. The MOSDEF survey used the MOSFIRE multi-object NIR spectrograph \citep{mclean2012} on the Keck I telescope to observe galaxies and AGN at $z\sim$1.37 -- 3.80 over the course of four years; for this study we use the data from the first three years of the survey. Our aim is to resolve the outstanding question of whether the identification of \irx sources is due to an overestimation of the IR-based SFR, an underestimation of the UV-based SFR, and/or AGN contamination.

With the MOSDEF survey, unlike previous studies of \irx galaxies, we are not limited to BzK galaxies and benefit from spectroscopic measurements for \z galaxies. We investigate how the identification of \irxg depends on the choice of templates used for estimating \lir and the methods used for the reddening correction of $\rm SFR_{UV}$. The spectroscopic data in MOSDEF allows us to investigate whether other SFR tracers such as \ha \; are elevated in \irx galaxies compared to the normal galaxies.
As discussed above, for identifying the contribution due to AGN previous studies had to rely only on X-ray stacking analyses. In this study we additionally use the rest-frame optical spectra obtained in the MOSDEF survey and the BPT diagram \citep[e.g.][]{baldwin1981}, as well as MIR identification of AGN, to investigate the contribution of AGN to the observed MIR-excess.

The outline of the paper is as follows: in Section \ref{sec:data} we describe our dataset, the procedure used for measuring the optical emission lines, the 24 \um and FIR data, the IR stacking analysis, and the method used for estimating stellar masses. In Section \ref{sec:results} we present our results on the nature of obscured AGN at $z\sim2$. In this section we consider various methods for estimating \sfir and \sfuv and extinction correction of $\rm SFR_{UV}$. We additionally investigate whether AGN can result in MIR-excess and perform X-ray stacking analysis. We discuss our results in Section \ref{sec:discussion} and conclude in Section \ref{sec:summary}. Throughout the paper we adopt a flat cosmology with $\Omega_{\Lambda}$ =0.7 and $H_0$=72 km s$^{-1}$Mpc$^{-1}$.
We assume a \citet{chabrier2003} stellar initial mass function (IMF) unless stated  otherwise.

\section{Data}
\label{sec:data}

In this paper we use data from the first three years of the MOSDEF survey to study the nature of \irx galaxies at \zz. In this section, we describe our galaxy and AGN sample and the methods used for measuring emission lines and stellar masses. We also describe the 24 \um and FIR data used for our sample, as well as the IR stacking analysis that we perform.


\subsection{The MOSDEF Survey}
\label{sec:mosdef}

For the MOSDEF survey \citep{Kriek2015} we obtained spectroscopic data from the MOSFIRE spectrograph \citep{mclean2012} on the 10 m Keck I telescope. MOSFIRE provides spectroscopic data with wavelength coverage from 0.97 to 2.40 \um\; at moderate resolution (R=3000-3650). 
\rf{We obtained spectra for sources in five extragalactic fields: AEGIS, COSMOS, GOODS-N, GOODS-S, and UDS. MOSDEF targets were selected from the deep $H$-band imaging provided by the CANDELS survey \citep{grogin2011,koekemoer2011} in areas with overlapping coverage from the 3D-\textit{HST} grism survey \citep {brammer2012,Momcheva2016}.}
The targets were selected in three redshift intervals ($1.37<z< 1.70$, $2.09 < z<2.61$ and $2.95 < z <3.80$)  to ensure coverage of the strong rest-frame optical emission lines (i.e., \oii, \hb, \ox, \ha, \nii, \sii) in the Y, J, H, and K bands. AGN were identified prior to targeting using X-ray imaging from \textit{Chandra} and/or IR imaging from IRAC on the \textit{Spitzer} telescope. AGN were given higher targeting weights, as were brighter sources and sources with more secure prior redshift estimates, either photometric or spectroscopic.
For the full details of the MOSDEF survey see \cite{Kriek2015}.

Our full sample using data from the first three years of the MOSDEF survey includes 792 galaxies, 92 of which are identified as AGN. For this study we limit our sample to sources at $1.4<z<2.61$ to ensure that 24 \um\; observations from MIPS on \textit{Spitzer} correspond to rest-frame $\sim$8 \um data. We further restrict our analysis to sources with significant 24 \um and UV detections, which limits our sample in this paper to 194 galaxies, 52 of which are AGN (see Section \ref{sec:agn}).

\subsection{Emission Line Measurements}
\label{sec:spectra}

Using the MOSDEF spectroscopy, we measure the flux of multiple rest-frame optical emission lines and use this to estimate the Balmer decrement (from \ha/\hb) and to identify optical AGN using the \ybpt\  versus  \xbpt\ BPT diagram. In this section we briefly describe the methods used by \cite{reddy2015} and \cite{Azadi2017} to  measure the flux of these emission lines. 

For MOSDEF galaxies the \hb, \ox, \ha, and \nii\ emission line fluxes are estimated by fitting Gaussian functions as described in \cite{reddy2015}. We allow a linear fit to the continuum, a single Gaussian component for \hb\ and \ox , and a triple  Gaussian function for \blun, \ha\ and \redn\, fit simultaneously. The spectra of each galaxy were perturbed 1000 times within the error spectra and the dispersion of the perturbed lines was taken as an estimate of the error in the emission line flux \citep{reddy2015}.

To identify optical AGN, we use the  procedure described in \cite{Azadi2017}. We use the \texttt{MPFIT} nonlinear least-square fitting function in IDL \citep{markwardt2009non} and use the error spectra to determine the errors on the fit. 
We require the continuum around each emission line to be flat, and we require the same physical components (i.e., narrow line, broad line, outflow) to have the same FWHM and velocity offset, as determined from the line with the highest S/N. The spacing between the  \bluox\; and \redox\; (as well as the \blun\; and \redn\ lines) forbidden lines are fixed and their flux ratios are set to be 1:3. We fit  \redox\ simultaneously with \bluox, and \ha\; with \blun\; and \redn\; using four different models. 

In model 1, all lines are fit with a narrow component with FWHM $<$2000 km s$^{-1}$. In model 2, in addition to the narrow component we allow for an underlying broad component for the \hb\; and \ha\; lines with FWHM $>$2000 km s$^{-1}$. In model 3, we fit each line with a narrow component and an additional component with FWHM $<$2000 km s$^{-1}$ and a velocity offset relative to the narrow component; this represents an outflow component. In model 4, the narrow, outflow, and broad \hb\; and \ha\; components are all fit. The fit from model 1, with only single narrow components, is considered to be the best fit for the sources unless the additional components improve the reduced $\chi^{2}$ at the 99\% confidence level. For more details on the AGN line fitting procedure see Section 2 in \cite{Azadi2017}. We note that the flux of the narrow components of the \hb\; and \ha\; lines are corrected for the underlying stellar absorption \citep[for more details see][]{reddy2015}. 

\raggedbottom

\subsection{AGN Sample}
\label{sec:agn}
In the MOSDEF survey we use three methods to identify AGN: X-ray imaging from \textit{Chandra}, IRAC MIR imaging from \textit{Spitzer}, and the BPT diagram using the MOSDEF spectral line measurements.  Details of MOSDEF AGN identification using each method are described in \cite{Azadi2017}; relevant details are repeated here. 

X-ray AGN were identified prior to MOSDEF targeting using  \textit{Chandra} observations with a depth of 4 Ms in GOODS-S, 2 Ms in GOODS-N, 800 ks in AEGIS, and 160 ks in the COSMOS field, corresponding to hard band \rf{flux ($f_\mathrm{2-10\ keV}$ [erg~$\rm s^{-1}$~cm$^{-2}$]) limits} of $1.6 \times 10^{-16}$, $2.8 \times 10^{-16}$, $5.0 \times 10^{-16}$, and $1.8 \times 10^{-15}$, respectively, in each field.  X-ray catalogs were generated according to the method described by \cite{laird2009}, \cite{nandra2015} and \cite{Aird2015}. To identify the optical, NIR and \textit{Spitzer} IRAC counterparts of the X-ray sources we use the likelihood ratio technique \citep[e.g.][]{Brusa2007,laird2009,Aird2015} where the source with the highest likelihood ratio is selected as the best match for sources with multiple counterparts. We then matched the X-ray counterparts to the 3D-\textit{HST} catalogs used for targeting in the MOSDEF survey.  We estimate the rest-frame 2--10 keV X-ray luminosities based on the hard (2--7 keV) or the soft (0.5-2 keV) band flux (when there is no detection for the hard band), assuming a simple power-law spectrum with Galactic absorption, photon index of $\Gamma$ = 1.9 and the MOSDEF redshift.

Although X-ray imaging is reliable for AGN identification, it may fail to identify highly obscured and Compton-thick AGN. IR imaging can recover some of these heavily obscured AGN, as UV and optical photons from the central engine are absorbed by dust grains in the obscuring structure and re-emitted at longer wavelengths. Several mid-IR AGN selection techniques have been proposed using IRAC \citep{fazio2004} data from the \textit{Spitzer} telescope. In MOSDEF we use the \cite{donley2012} criteria with some slight modifications. The \cite{donley2012} criteria are:
  
 \begin{eqnarray}
  x={\rm log_{10}}\left( \frac{f_{\rm 5.8 \um}}{f_{\rm 3.6 \um}} \right), \quad
  y={\rm log_{10}}\left( \frac{f_{\rm 8.0 \um}}{f_{\rm 4.5 \um}} \right)
\end{eqnarray}

\begin{align}
  &x\ge 0.08 \textrm{~ and ~} y \ge 0.15 \textrm{~ and ~} \label{eq2}  \\ 
  &y\ge (1.21\times{x})-0.27 \textrm{~ and ~} \label{eq3}  \\
  &y\le (1.21\times{x})+0.27 \textrm{~ and ~} \label{eq4} \\
  &f_{\rm 4.5 \um} > f_{\rm 3.6 \um} \textrm{~ and ~}  \label{eq5} \\
  &f_{\rm 5.8 \um} >  f_{\rm 4.5 \um} \textrm{~ and ~} \label{eq6}  \\
  &f_{\rm 8.0 \um} > f_{\rm 5.8 \um}. \label{eq7} 
\end{align}

We require detections in all IRAC bands with S/N limits of 3, 3, 2.4, and 2.1, respectively in channels 1 to 4 \citep{mendez2013primus}. 
Our slight modifications are that we do not require the criteria listed in equation \ref{eq4} above and we allow equations \ref{eq5}, \ref{eq6} and \ref{eq7} to hold within the 1$\sigma$ errors on the IRAC photometry. The \cite{donley2012} criteria are very pure, which allows for the creation of a sample without contamination from non-AGN, but are also fairly restrictive. These slight modifications are made to include a handful of additional sources that are highly likely to be AGN \citep[for more details see][] {Azadi2017}.

We also use the BPT digram to identify optical AGN. To measure the \xbpt\; and \ybpt\; line ratios we use the line fitting procedure described in Section \ref{sec:spectra}. We  exclude any sources that require a significant broad component for either of the \hb\; or \ha\; lines since broad emission lines can dominate over the host galaxy in the broad band photometry and affect the stellar mass measurements. As shown in \cite{coil2015} and \cite{Azadi2017}, we consider sources above the  \cite{melendez2014} line to be optical AGN in our sample, as this line provides a reliable AGN sample for MOSDEF sources when compared to X-ray and IR AGN in our sample.


\subsection{Stellar Mass Measurements}
\label{sec:mass}

We estimate stellar masses in MOSDEF using spectral energy distribution (SED) fitting of the multi-wavelength photometry from 3D-\textit{HST} \citep{skelton2014} and our spectroscopic redshift measurements. We use the FAST stellar population fitting code \citep{kriek2009}, with a \cite{chabrier2003} stellar initial mass function (IMF), \cite{conroy2009} Flexible Stellar Population Synthesis (FSPS) models, the \cite{calzetti2000} dust reddening curve, and a fixed solar metallicity, with delayed exponentially-declining star formation histories. 

Radiation from AGN may contribute to the observed SED, particularly at UV and MIR wavelengths, which could impact our derived stellar masses of their host galaxies. 
We previously tested fitting the potential contribution from AGN to the SED using additional power-law contributions to the photometry at UV and MIR wavelengths \citep{Azadi2017}. The FAST code can be run with or without the 
template error function \citep[which accounts for wavelength-dependent mismatch between the observed photometry and the templates, e.g.][]{kriek2009}, and as shown in \citet{Azadi2017} the stellar masses that we obtain for MOSDEF AGN by running FAST  without the template error function and subtracting power-law components for the AGN light are consistent with those estimated using the template error function (and not subtracting additional AGN components). Therefore in this study we use results from FAST with the template error function.


\subsection{24 \um Flux Density and $L_{8}$ Measurements}
\label{sec:l8}

In this study we restrict our analysis to galaxies with $3\sigma$ \textit{Spitzer}/MIPS 24 \um detections.
We use 24 \um data from the Far-Infrared Deep Extragalactic Legacy (FIDEL) Survey \citep[]{Dickinson2007}. We use the measurements of \cite{Shivaei2017} for the 24 \um flux densities. 
In brief, \citeauthor{Shivaei2017} \citep[also see][]{Reddy2010} use higher spatial resolution data from IRAC to determine the location of each object. Around each object a sub-image is constructed and a point spread function (PSF) is fit simultaneously to all sources, including the object of interest, and a covariance matrix is used to determine the robustness of each fit. The background flux is estimated by fitting the PSF to random positions at least 1 FWHM away from the object, and the standard deviation of those fluxes is taken as the noise.

As noted above, we limit our sample to galaxies and AGN  with $1.4<z<2.61$ to ensure that their 24 \um\; flux densities cover the rest-frame $\sim$ 8 \um luminosity ($\rm L_{\rm 8} = \nu L_{\nu}[8\mum]$). In our analysis we use various luminosity-dependent template sets \citep[e.g.][]{CE01,DH02,Rieke2009} to extrapolate the total IR luminosity (\lumir) from the observed 24 \um band; we describe these templates and the procedure to estimate \lir fully in Section \ref{ir1} below.
We estimate the rest-frame
8 \um luminosity, $\rm L_{\rm 8}$, based on the same template. The values of \leight estimated in this manner do not depend significantly on which template
set is used; results using the 
\citeauthor{CE01}, \citeauthor{DH02} or \citeauthor{Rieke2009} template sets are consistent with each other.


\subsection{\textit{Herschel}/PACS Data and IR Stacking Analysis} \label{sec:herschel}

As we discuss below in Section \ref{sec:results}, including FIR data plays an important role in accurately estimating the bolometric IR luminosity which allows us to more cleanly select galaxies where the IR emission is substantially elevated relative to the dust-corrected UV.
In part of our analysis below (see Section  \ref{ir3}) we include the \textit{Herschel}/PACS 100 and 160 \um data available in the AEGIS, COSMOS, and GOODS  fields \citep{Oliver2010,Elbaz2011,Magnelli2013}. The number of  MOSDEF galaxies with robust PACS 100 and 160 \um detections is 63 and 58, respectively (40 galaxies have detections at both wavelengths). To measure the 100 \um and 160 \um flux densities we use the prescription described above for 24 \um flux densities. \rf{We note that the majority of the MOSDEF galaxies do not have a robust detection at 24 \um or FIR wavelengths. While this study is limited to sources with significant 24 \um detection, the \lir from \cite{Shivaei2017} used in section \ref{ir3} is estimated using stacks of the MIPS and Herschel images}.


\section{Analysis and Results}
\label{sec:results}

In this section we first define \irx galaxies (Section~\ref{sec:defirx}) \rf{and then describe the various methods used in this study for estimating bolometric infrared luminosities and correcting the UV SFR for dust absorption in Section 3.2. In Section ~\ref{sec:irxinmosdef} we identify \irx galaxies in MOSDEF using these different methodologies.}
We investigate whether the incidence of \irx depends on stellar mass (Section \ref{sec:stellarmass}) and whether detected or obscured AGN contribute to the identification of \irx galaxies (Sections \ref{sec:detectedagn} and \ref{sec:stacks}).
We further consider whether the SFR traced by  H$\alpha$ is elevated in \irx compared to normal galaxies (Section \ref{sec:sfha}).


\subsection{Definition of MIR-Excess Galaxies}
\label{sec:defirx}

Studying a sample of BzK-selected galaxies at \z, \da report the existence of a population of so-called ``MIR-excess" galaxies for which the total IR luminosity as  estimated from the MIPS 24 \um flux densities shows an excess relative to that expected from other SFR tracers. Specifically, they identify \irx sources as galaxies satisfying the following criterion:
\setlength{\abovedisplayskip}{10pt}
 \begin{eqnarray}
{\rm log}\left( \frac{\rm SFR_{\rm IR}+ \rm SFR_{\rm UV}}{\rm SFR_{\rm UV,corr}} \right) > 0.5
\end{eqnarray}
where \sfir is the SFR estimated from the MIPS 24 \um flux density, \sfuv is the SFR estimated from the UV data, and \sfuvcorr is the SFR estimated from the UV data  corrected for dust extinction. 

The empirical threshold of 0.5 dex is chosen by \da assuming that the intrinsic scatter in this ratio is symmetric to first order.
In their study, \da use the luminosity-dependent \cite{CE01} templates to convert \rf{observed} 24 \um flux to $\rm L_{\rm IR}$. They correct \sfuv for dust reddening using the relation between the color excess and B-z color from \cite{Daddi2004}, which is derived using a  \cite{calzetti2000} attenuation law. \da find that $\sim 25\%$ of their BzK-selected galaxies at \z show such an excess at MIR wavelengths (after removing the X-ray detected AGN from their sample) and emphasize that the existence of this population is independent of the templates used for estimating the bolometric IR luminosity. 

As noted above, in this paper we aim to investigate how the templates used for estimating \lir and the choice of the dust correction methods used for estimating \sfuvcorr (which may depend on the adopted SFH as shown below) affect the identification of \irx galaxies.

\rf{\subsection{Methodology}}
\label{sec:all_methods}

\rf{In this section we describe various methods used in this study for estimating \lir and for correcting \sfuv for dust extinction. We implement these methods below in Section \ref{sec:irxinmosdef} in order to identify \irx galaxies in our sample.}

\subsubsection{Estimating the Total IR luminosity from Luminosity-Dependent Templates} \label{ir1}

\rf{The first sets of templates adopted in our analysis are luminosity-dependent templates from \cite{CE01}, \cite{DH02} and \cite{Rieke2009} which estimate \lir from a single 24 \um band observation. We describe each of these template sets in this section.}

\cite{CE01} construct their templates using data from \textit{ISO}, \textit{IRAS} and \textit{SCUBA} 
for local galaxies create 105 templates covering 0.1-1000 \um.\footnote{\rf{\cite{CE01} includes data from the ISOCAM-LW2 band centered at 6.7 \um which is broad enough to encompass the 7.7~\um PAH emission feature.}} 
\rf{The \lir (8--1000 \um) is determined using \cite{Sanders96} relation.}
These templates span a range of \lir of $10^{8}- 10^{13}$ \lsun, from normal galaxies to luminous and ultra luminous IR galaxies (LIRGS: $10^{11} $\lsun $\leqslant$\lir $\leqslant 10^{12} $\lsun and ULIRGS: \lir $ \geqslant 10^{12} $$\rm L_{{\odot} }$). The templates are luminosity-dependent: a single template describing the IR emission is specified for a given luminosity.

To estimate \lir using the \citeauthor{CE01} templates, we  first shift the templates to the redshift of each source and then convolve them with the MIPS 24 \um response curve. From the 105 templates, we consider the one that at 24 \um returns the closest value to the observed flux as the best template and adopt the luminosity corresponding to that template from \cite{CE01} as the total IR luminosity.
We convert \lir to \sfir using the relationship given in \cite{kennicutt1998star}, converted to a \cite{chabrier2003} IMF. Thus,
\begin{equation}
{\rm{SFR_{IR}} [M_\odot \mathrm{yr}^{-1}]= 2.8 \times 10^{-44} \times {\rm L_{IR (8-1000\; \mum)}} [\rm{erg\;s^{-1}}].} 
\label{eq:lirtosfr}
\end{equation}



The semi-empirical templates of \cite{DH02} are constructed using data from \textit{ISO}, \textit{IRAS} and \textit{SCUBA} for 69 local galaxies with ${\rm L_{IR(8-1000\; \mu m)}}$ $< 10^{11}$ $\rm L_{{\odot} }$. \cite{DH02} develop a theoretical model with three components for the emission from small and large dust grains, as well as PAH molecules \citep[see also][]{Dale2001}. In their model the mass of each dust grain has a power-law distribution as a function of intensity with a slope of $\alpha$ and different templates are constructed based on variations of $\alpha$ from 1--2.5.
We use \cite{DH02} templates with the empirical calibration of \cite{Marcillac2006} to scale each template (for a given $\alpha$) to a specific total \lir \citep[see Equation 1 in][also see Papovich et al. 2007 and Reddy et al. 2012]{Marcillac2006}.
We integrate these templates to estimate the total \lir and find the template that provides the best match to the observed 24 \um flux in a similar fashion to that described above for the \cite{CE01} templates.
We convert \lir to \sfir using Equation~\ref{eq:lirtosfr}.


\cite{Rieke2009} consider a sample of 11 local LIRGS and ULIRGS with \textit{Spitzer}/IRS spectra and additional photometric data from optical to radio wavelengths. \cite{Rieke2009} calculate the IR luminosity for each of their galaxies using the \cite{Sanders2003} calibration  and combine the SED of the galaxies with similar luminosities to build a set of average templates with  ${\rm L_{IR(5-1000\; \mu m)}}$ between $10^{9.75} - 10^{13}$ \lsun.
We estimate \sfir directly from the 24~\um flux densities using equation 14 in \citet{Rieke2009}. We note that \cite{Rieke2009} assumes 
 a \cite{Rieke1993} IMF, which according to the author has a similar behavior to the \citeauthor{chabrier2003} IMF assumed throughout this paper.


\subsubsection{Estimating the Total IR Luminosity from Luminosity-Independent Conversions} \label{ir2}

The luminosity-dependent templates described above are created based on local galaxies and may not be applicable at higher redshifts. With the advent of \textit{Herschel}, direct detection of individual \z galaxies at FIR wavelengths has become possible. Studies show that extrapolations from \rf{observed} 24 \um can result in an overestimation of \lir in luminous galaxies at \z \citep[e.g.][]{Nordon2010,Elbaz2011,Reddy2012}, as these galaxies may have a different dust temperatures and different luminosity surface densities than \e{luminous local galaxies} \citep[e.g.][]{Symeonidis2011,Rujopakarn2013}. 
\rf{We also note that at $z\sim2$ the observed 24 ~\um band is probing PAH emission regions. 
The intensity of the PAH emission can vary significantly and the shape of the SED is thus poorly constrained at these wavelengths.}

While in the local Universe there is a positive correlation between galaxy luminosity and dust temperature, at \z there is a large scatter in their relation \citep[e.g.][]{Magdis2010,Elbaz2010,Symeonidis2011,Wuyts2011}. Using deep 100 to 500 \um FIR data from \textit{Herschel}, \cite{Elbaz2011} show that at $z>1.5$ a \irx can occur in individual galaxies with \lir $> 10^{12}$~\lsun and argue that while local LIRGS and ULIRGS are typically in a starburst mode, at \z they are typically on the star-forming main sequence and therefore can be treated as scaled up versions of local main sequence galaxies \citep[e.g.][]{Elbaz2011,Nordon2012}.

To address the problem of overestimation of \lir for \z ULIRGS, \cite{Elbaz2011} propose a single luminosity-independent conversion \citep[see also ][] {Wuyts2008,Wuyts2011,Nordon2012}. They use a normalization derived from the \cite{CE01} templates, identify the best template as the one that best fits the \textit{Herschel} data, and calculate \lir by integrating over the best fit SED. 
\cite{Elbaz2011} find that the bolometric correction factor at 8 \um ($\rm L_{IR}$/\leight) corresponds to a universal value of 4.9 at $z<2.5$.

\rf{Another commonly used luminosity-independent template is provided by \cite{Wuyts2008}.} \cite{Wuyts2008} use the \cite{DH02} models and find $\rm L_{IR,\alpha}$ corresponding to different $\alpha$ values. They then use the average  $\rm L_{IR,\alpha}$ as the best luminosity for each galaxy. We note that unlike \cite{Elbaz2011}, the \cite{Wuyts2008} conversions are not determined using FIR data. In a later paper, \cite{Wuyts2011} show that the conversion factors from \cite{Wuyts2008} result in IR luminosities that are consistent in the median (although with a large scatter) with \textit{Herschel} \lir measurements. 

{\subsubsection{Estimating the Total IR Luminosity from Luminosity-Independent, Mass-Dependent Conversions} \label{ir3}

As shown in \cite{Elbaz2011}, a luminosity-independent bolometric correction factor, $\rm L_{IR}$/\leight, 
calibrated using \textit{Herschel} FIR data,
can provide a robust estimate of \lir for galaxies. 
However, it has been shown that due to the variations of PAH intensity with metallicity the ratio of $\rm L_{IR}$/\leight varies  with stellar mass, metallicity, and/or sSFR \citep[e.g.][]{Nordon2012,Shivaei2017}. In this section we test the mass-dependent, luminosity-independent bolometric factor of \cite{Shivaei2017} to estimate $\rm L_{\rm IR}$. 

As noted in Section \ref{sec:herschel}, the majority of MOSDEF galaxies are not detected in the \textit{Herschel}/PACS bands. To derive \reight ratios for the MOSDEF galaxies, \cite{Shivaei2017} stack \rf{MIPS 24 \um and \textit{Herschel}/PACS 100 and 160 \um images of MOSDEF sources regardless of their detection} and fit the stacks with the luminosity-dependent templates of \cite{CE01}. They estimate the average infrared luminosity over the 8-1000 \um range, $\rm <$$\rm L_{IR}$$\rm>$, using the best fit template to the stacked data. 
\cite{Shivaei2017} repeat the same prescription for estimating $\rm<$$\rm L_{8}$$\rm>$ and the ratio \reight as a function of stellar mass. The \cite{Shivaei2017} method is based on the assumption that the luminosity scaling of the \cite{CE01} templates applies to \z galaxies and that these templates can accurately estimate \lir when fitted to only the PACS bands. We note that the calculations in \citet{Shivaei2017} are at 7.7 \um, here we adopt the correction factor of $\rm L_{7.7}$/\leight = 1.25 reported in their paper for our analysis. \rf{We use this method below, along with the other methods discussed above, in Section \ref{sec:irxinmosdef} to estimate \lumir.}


\subsubsection{Correcting \sfuv for Dust Absorption} 
\label{uvcorr}

\rf{In this section we describe how the \sfuv is estimated in our work and describe the four methods used to correct \sfuv for dust extinction.}
To estimate \sfuv (uncorrected for dust extinction), we fit the SEDs of our sources using the 3D-\textit{HST} photometric catalogs and measure the UV flux (at a rest-frame wavelength 1600 \AA) from the best fit stellar population model \citep[for more details  see][]{reddy2015}. We convert the estimated luminosity to \sfuv using the \cite{kennicutt1998star} relation adjusted to a \cite{chabrier2003} IMF:
\begin{equation}
{\rm{SFR_{UV}} [M_\odot {yr}^{-1}]= 7.7 \times 10^{-29}\times L_{\nu (1600 \AA)} [erg s^{-1} Hz^{-1}]}
\label{eq:luvtosfr}
\end{equation}
The dust-corrected \sfuvcorr is given by \sfuvcorr$=10^{0.4\rm A_{\rm UV}} \times \rm SFR_{UV}$, where $A_{\rm UV}$ is the UV extinction.
To estimate $\rm A_{UV}$ we either use the color-excess, E(B-V), from SED fitting assuming $\rm A_{\lambda}$ = K($\lambda$) E(B-V) where K is the  attenuation curve, or use the UV continuum slope. 
We also investigate the effect of changing the assumed dust attenuation law and the assumed SFH in the SED modeling. 
For the dust attenuation we compare the widely-used \citet{calzetti2000} law with the more recent attenuation curve derived by \citet{reddy2015} using data from the first two years of the MOSDEF survey.  
For the SFH, we investigate two models:
i)~a delayed exponentially-declining SFH, and ii)~a constant SFH. We investigate the effect of the adopted dust attenuation law and SFH on the UV extinction correction and consequently on our estimates of \sfuvcorr and the fraction of \irx sources that are identified.


In our first method for correcting $\rm SFR_{UV}$,
we measure E(B-V) by fitting the SEDs using the FAST code \citep{kriek2009} with a \cite{chabrier2003} IMF, \cite{conroy2009} FSPS models, a delayed exponentially-declining SFH with $\tau$ varying from 100 Myr to 1 Gyr, \cite{calzetti2000} reddening curve, and assuming a fixed solar metallicity. In the second method we fit the SEDs with all of the same input parameters except we use the \cite{reddy2015} attenuation curve.

In the third method, we use the UV spectral slope, $\beta$, to estimate the dust correction and derive \sfuvcorr for each galaxy in our sample. 
Rather than measuring the slope directly from the photometry, we measure the slope from the fits to the SEDs (we discuss this choice below in Section \ref{discuss1}). The input parameters in the SED fitting remain the same as in the row above.
To measure the UV slopes we follow the prescription of \cite{Calzetti1994} and use several continuum windows spanning 1250--2600 \AA \; to avoid the stellar absorption features at these wavelengths. We then find the \emph{average} relation between $A_{\rm UV}$, as recovered from the full SED fits to each of our galaxies, and the observed value of $\beta$:
\begin{equation}\label{eq:auvbeta}
\rm A_{\rm UV}= 1.99 \beta + 4.13  
\end{equation}
Finally, we re-calculate $A_{\rm UV}$ for each \emph{individual} galaxy based on the measured $\beta$ and the average relationship given in Equation~\ref{eq:auvbeta} above. 


We note that our derivation of the average relationship between $A_{\rm UV}$ and $\beta$ given in Equation \ref{eq:auvbeta} depends on the assumed dust attenuation law and SFH. Our final method for measuring $\rm A_{UV}$ is
using the relation between $A_{\rm UV}$ and $\beta$ from \cite{reddy2015}:
\begin{equation}\label{eq13}
\rm A_{UV}= 1.84 \beta + 4.48  
\end{equation} 
\cite{reddy2015} obtain this relation using the same method described above for 
Equation \ref{eq:auvbeta} but instead of a delayed exponentially-declining SFH they assume a constant SFH with 
a fixed age of
100 Myr when performing the initial SED fitting. 
Under this assumption, a higher amount of dust is required to turn an intrinsically blue SED into a given observed SED, while with a slowly declining SFR both populations of old and newly formed stars exist. This mix of stellar population can match an observed SED without requiring as much dust \citep[e.g.][]{Schaerer2013}.
Additionally, we note that assuming a constant SFH with an age that is allowed to vary over a range similar to the e-folding time in our delayed-$\tau$ model (100 Myr -- 1 Gyr), results in a similar \sfuvcorr as in our third method.

\begin{figure*}
\includegraphics[width=15cm,center]{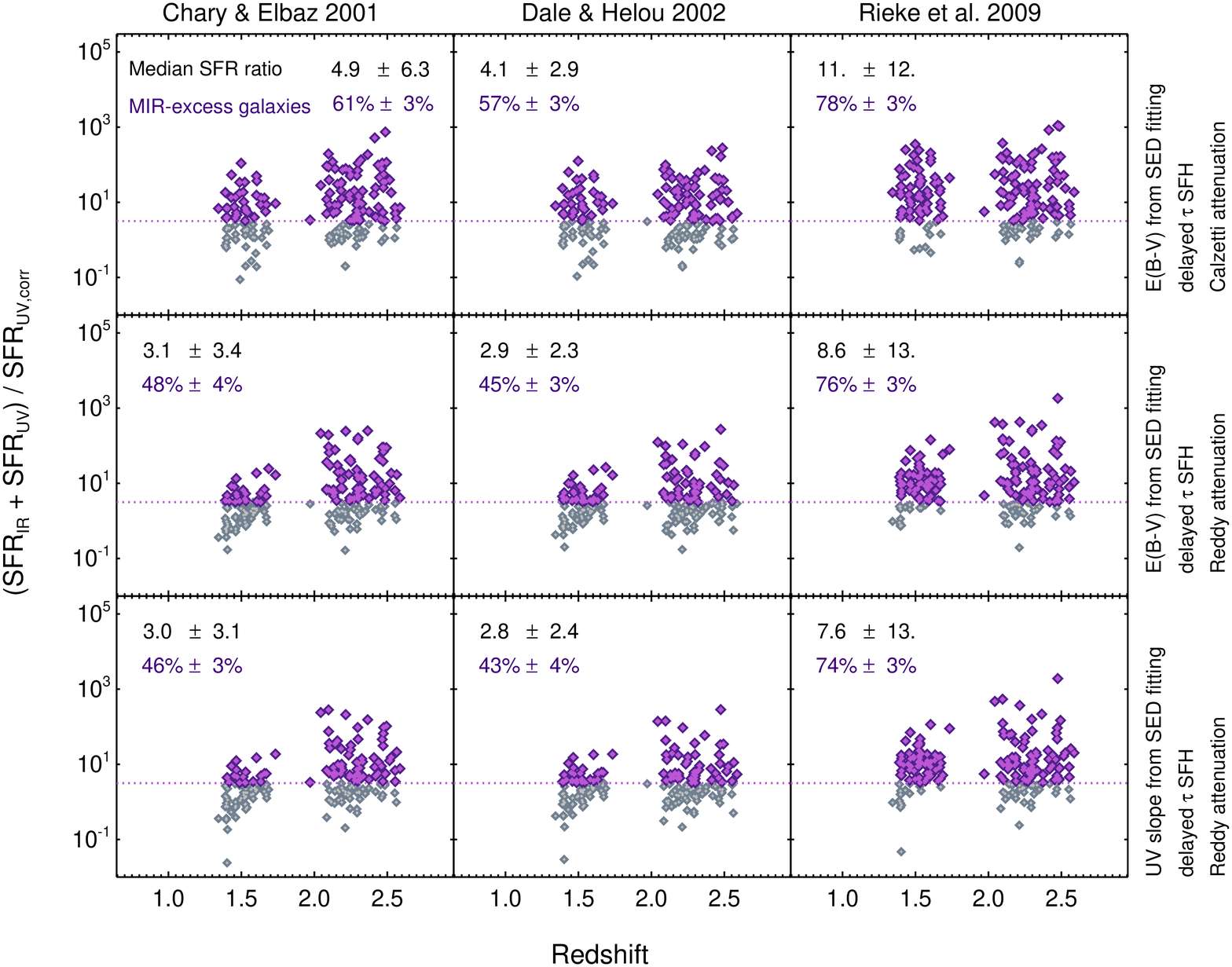}
\caption{\e{\sfratio versus redshift for MOSDEF galaxies with significant 24 \um detections. The dotted purple line shows \sfratio $\sim$ ~3.2 which is the threshold considered by \cite{Daddi2007a} for \irx galaxies identification. The bolometric IR luminosity is estimated by extrapolating from the 24 \um band using the luminosity-dependent templates of \cite{CE01}, \cite{DH02} and \cite{Rieke2009} in each column. Each row corresponds to a different method for estimating the extinction correction to \sfuv. The methods used, starting at the top row, are: estimating the color-excess from the SED fitting with a delayed-$\tau$ SFH and assuming the \cite{calzetti2000} attenuation curve and \citet{reddy2015} attenuation curve, and using UV spectral slope as measured by fitting power-laws at 1200-2600\AA\ to the best SED fit. The percentage  of 24 \um-detected  galaxies that are also \irx galaxies is shown in purple text, and the black text shows the median \sfratio in each panel. The fraction of \irx galaxies identified depends on the templates used to estimate \lir and the extinction correction method.}}
\label{fig:lumdep}
\end{figure*}

\begin{figure*}
\includegraphics[width=15cm,center]{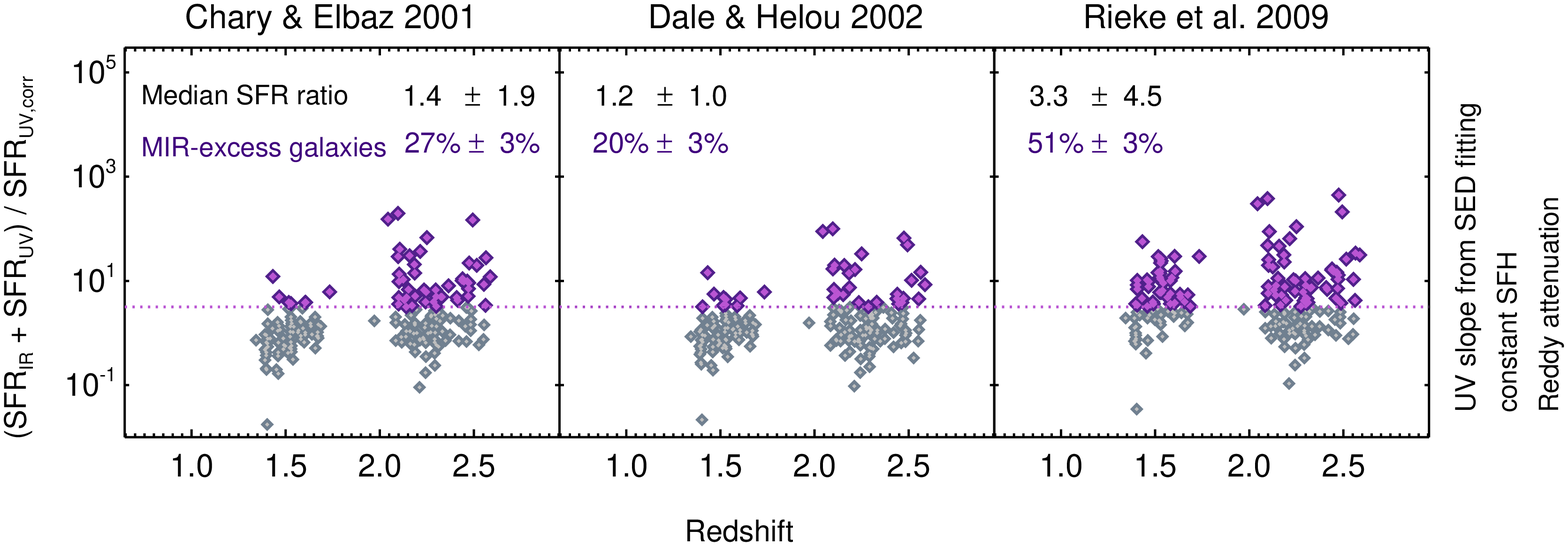}
\caption {\e{\sfratio versus redshift for MOSDEF galaxies with significant 24 \um detections. The bolometric IR luminosity is estimated by extrapolating from the 24 \um band using the luminosity-dependent templates of Figure \ref{fig:lumdep}. For estimating the \sfuv extinction correction, we measure the UV spectral slope from by fitting power-laws at 1200-2600\AA\ to the best SED fit assuming a constant SFH model with $\tau$ = 100 Myr (rather than a delayed $\tau$ model considered in Figure \ref{fig:lumdep}). The fraction of galaxies identified as \irx drops significantly compared to the last panel of Figure \ref{fig:lumdep}.}}
\label{fig:sfh}
\end{figure*}

\begin{figure}
\includegraphics[width=0.60\textwidth,center]{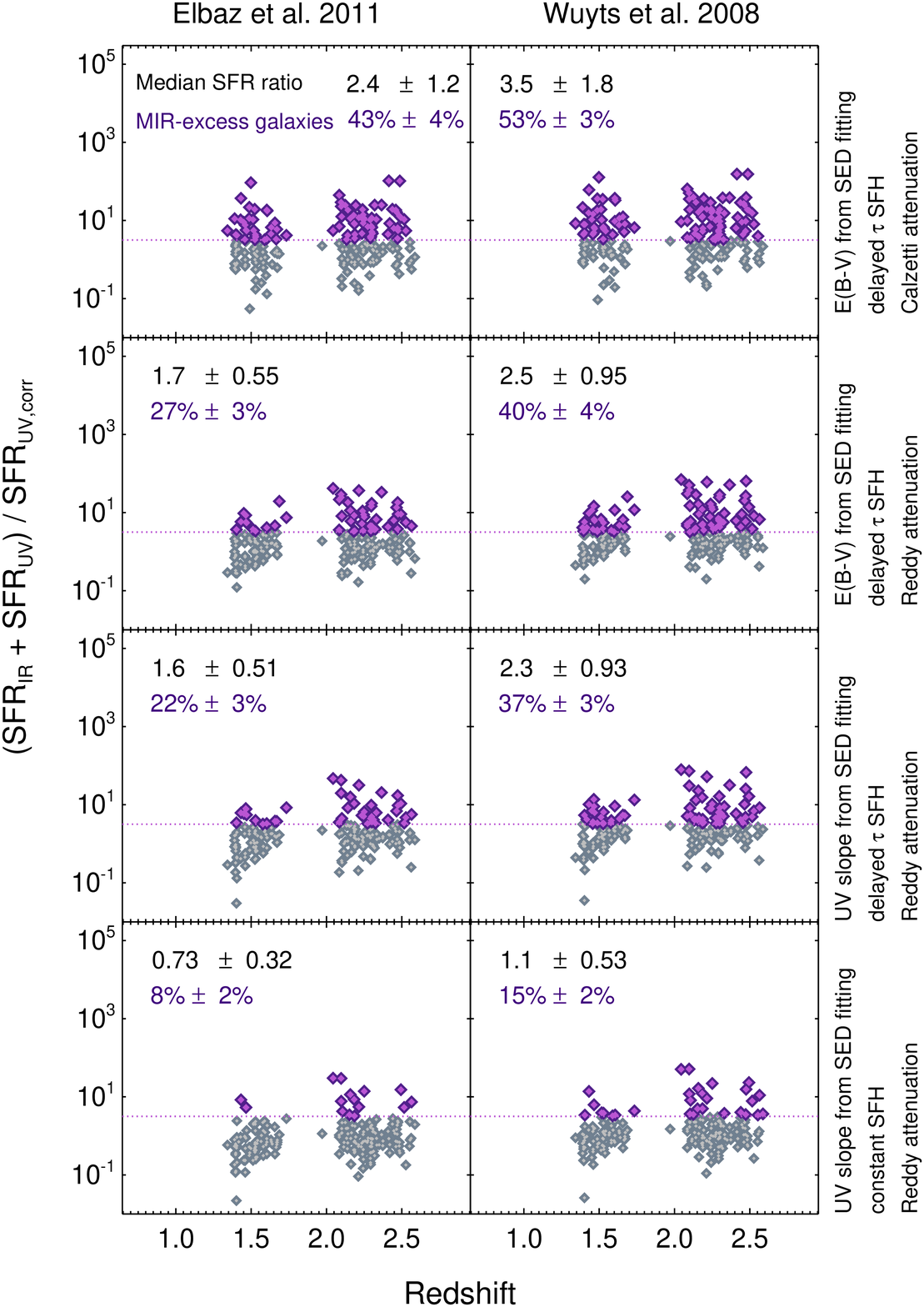}
\caption{\sfratio versus redshift for MOSDEF galaxies with significant 24 \um detections. \e{The dotted purple line shows \sfratio $\sim$ 3.2 which is the threshold considered by \cite{Daddi2007a} for \irx galaxies identification.} In the left column the total IR luminosity is estimated from \cite{Elbaz2011}, using a universal ratio of $\rm L_{IR}$/\leight = 4.9 which was determined by including \textit{Herschel} FIR data. In the right column the luminosity-independent, redshift-dependent conversions of \cite{Wuyts2008} derived using 24 \um data are used for estimating $\rm L_{\rm IR}$. Each row corresponds to a different method used for extinction correction of \sfuv and the methods are the same as in Figure \ref{fig:lumdep} and \ref{fig:sfh}. As in Figure \ref{fig:lumdep}, we report in each panel the fraction of 24 \um-detected galaxies that are MIR-excess, as well as the median $(\rm SFR_{\rm IR}+SFR_{\rm UV})/\rm SFR_{ \rm UV,corr}$. The fractions of \irx identified galaxies are lower in this figure than in Figure \ref{fig:lumdep} and \ref{fig:sfh}, for a given extinction correction method, and are lower here in the left column when FIR data are included when estimating $\rm L_{\rm IR}$.}
\label{fig:lumindep}
\end{figure}

\begin{figure}
\includegraphics[width=0.40\textwidth \hspace{2em}]{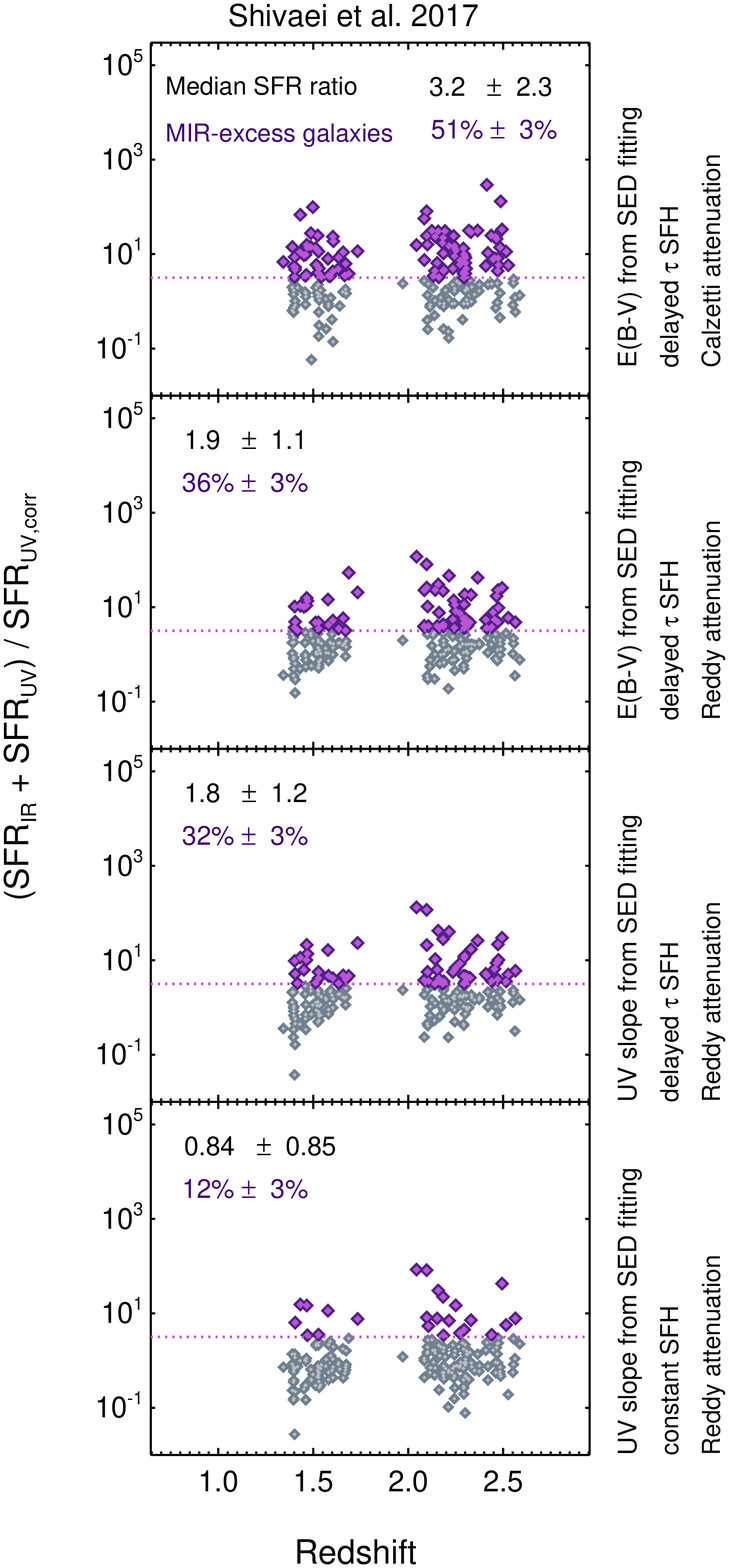}
\caption{\sfratio versus redshift for galaxies with significant 24 \um detections, determined using the the luminosity-independent, mass-dependent \reight from \cite{Shivaei2017}.\sfuv extinction correction methods are as in those described in Figure \ref{fig:lumdep} and \ref{fig:sfh}}. Here again we report the  fraction  of 24 \um-detected  galaxies that are \irx, as well as the median \sfratio. Overall, the fraction of \irxg and the median \sfratio drop significantly compared to the values in the left column of Figure \ref{fig:lumdep}.
\label{fig:massdep}
\end{figure}

Another approach for estimating the UV continuum reddening is using the nebular reddening. In general, it has been shown that the visual extinction of nebular lines is greater than the reddening of the UV-optical continuum, as the ionizing radiation originates from regions close to the dusty molecular clouds \citep[e.g.][]{Calzetti1994,calzetti2000}. We additionally tested correcting the UV emission for dust absorption and estimating the \sfuvcorr and \sfratio using the ratio of the Balmer emission lines (\ha/\hb) and the relation between the nebular and stellar E(B-V) from \citet[][equations 10 and 13]{reddy2015}. While we find that the fraction of \irxg identified with this \sfuvcorr is relatively low, this correction method results in substantial scatter in \sfratio compared to the other methods discussed above. This scatter is due both to the relatively low S/N in the observed \hb \ flux measurements and the large scatter around the average relation between nebular and stellar reddening \citep{reddy2015}.


\subsection{MIR-excess Galaxies in MOSDEF} 
\label{sec:irxinmosdef}

Figures \ref{fig:lumdep} to \ref{fig:massdep} illustrate the  \sfratio as a function of redshift for galaxies in the MOSDEF sample, where the \sfir in each column is estimated using the templates described in Sections \ref{ir1}, \ref{ir2} and \ref{ir3}, and \sfuv is corrected for extinction using different methods described in Section \ref{uvcorr}. In these figures we present the median \sfratio and the percentage of galaxies identified as \irx  in each panel, with errors estimated from bootstrap resampling. A median ratio close to one indicates a good agreement between \sfuvcorr and \sfuv$+$\sfir for that combination of methods. We note that in some galaxies \sfratio may legitimately be higher than one but we do not expect this ratio to be substantially greater than one for the bulk of the population. AGN are included in Figures \ref{fig:lumdep} to \ref{fig:massdep}; we discuss the impact of AGN on the \irx population in Section \ref{sec:detectedagn} below.

\begin{table*} 
\caption{The average bolometric correction factor, \reight, determined when stacked \textit{Herschel} data are included, using the methodology of  \cite{Shivaei2017}.}
\begin{center}
\renewcommand{\arraystretch}{2}
\begin{tabular} 
{c c c c}\hline \hline
\reight & Chary \& Elbaz (2001) & Dale \& Helou (2002)& Rieke et al. (2009)\\
\hline \hline
$\;\;\;9.6 <\log(\rm \dfrac{M_*}{M_\odot}) < 10.0$ & 13.9$\pm$3.1  & 12.5$\pm2.5$&15.6$\pm$3.9 
\\  
 $10.0<\log(\rm \dfrac{M_*}{M_\odot})<10.6$ & 5.21$\pm0.43$ & 4.63$\pm$0.34 &5.21$\pm0.43$\\ 
$10.6 <\log(\rm \dfrac{M_*}{M_\odot})<11.6$ &6.25$\pm$0.31  &4.81$\pm0.18$& 4.31$\pm0.15$  \\ 
\hline 
\end{tabular}
\end{center}
\label{tab:irxss3}
\end{table*}

We note that each of the methods used to calculate the SFR has uncertainties and caveats (e.g. the effect of the adopted SFH on the UV attenuation estimation, or the uncertainty on the UV continuum slope measurements in optically-thick galaxies). We discuss these issues further in Section \ref{sec:discussion} below.

Overall, Figures \ref{fig:lumdep} and \ref{fig:sfh} indicate that the percentage of 24~\um -detected galaxies classified as \irx galaxies can vary widely, from 20\%--78\%, depending on the templates used to estimate \lir and the \sfuvcorr estimation. In the first two rows of Figure \ref{fig:lumdep} the percentage of \irx galaxies drops due to a variation from \citeauthor{calzetti2000} attenuation curve to \citeauthor{reddy2015}. As discussed in \cite{reddy2015}, the overall shape of these attenuation curves is similar but the normalization of \citeauthor{calzetti2000} is higher.
The \citeauthor{reddy2015} curve thus results in slightly higher E(B-V) in the red galaxies which explains the lower percentage of \irx and median \sfratio in the second row of Figure \ref{fig:lumdep}. While the numbers do not change significantly from row 2 to 3 in Figure \ref{fig:lumdep}, when we adopt a constant SFH in Figure~\ref{fig:sfh} 
we find a significantly lower fraction of \irx galaxies.
However, the median ratio is closer to 1, indicating that using a constant SFH may result in a larger dust correction when estimating \sfuvcorr.

As mentioned above, when the ratio of \sfratio is close to unity the SFRs are in better agreement with each other, although each of the SFR tracers has uncertainties (see Section \ref{sec:discussion} which may result in a scatter around \sfratio $\sim 1$). The median \sfratio with the methods shown in Figure \ref{fig:lumdep} varies from 2.8 to 11.0. These high values are due primarily to an overestimation of \lir found when applying local templates at high redshifts, as we discussed above in Section \ref{ir2} and Section \ref{ir3}.

We note that using the \cite{CE01} templates and the \cite{calzetti2000} attenuation curve, \db find that $\sim 20\%-30\%$ of BzK-selected \z galaxies are \irx, while we find a much higher fraction of MOSDEF galaxies (61\%) with the same prescription. Below in Section \ref{sec:discussion} we discuss how various effects result in such a substantial difference in the incidence found in these studies. 

In Figure \ref{fig:lumindep} we use luminosity-independent conversions from \cite{Elbaz2011} and \cite{Wuyts2008} for estimating $\rm L_{IR}$. In the left column we use 
a universal ratio of $\rm L_{IR}$/\leight = 4.9 from \cite{Elbaz2011} and in the right column we use the conversion factors from \cite{Wuyts2008} for MIPS 24 \um fluxes at the redshift of each of our galaxies. Comparing Figures \ref{fig:lumindep} and \ref{fig:lumdep} indicates that using the \cite{Elbaz2011} method results in a substantial decrease in both the median \sfratio and the fraction of \irx galaxies compared to the \cite{CE01} method. However, given the errors, we do not find any significant  differences between the numbers resulting from the \cite{Wuyts2008} and the \cite{DH02} methods.  Overall, we find that 
a luminosity-independent conversion such as the  \cite{Elbaz2011} universal ratio leads to better agreement between $\rm SFR_{IR}+SFR_{UV}$ and $\rm SFR_{UV,corr}$ at $z\sim 2$.

\begin{figure}
\includegraphics[width=0.42\textwidth]{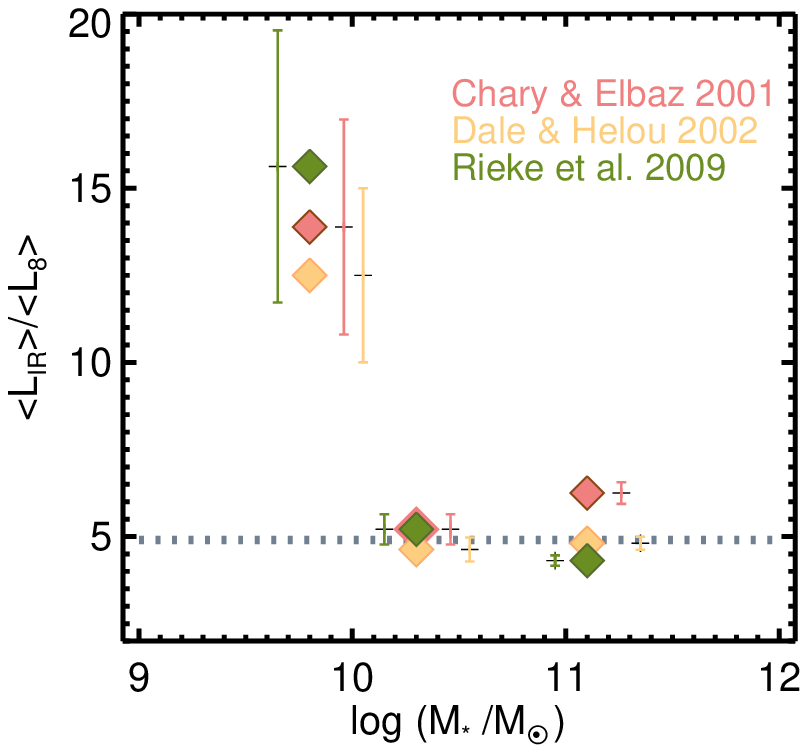}
\caption{The average bolometric correction factor, $\rm <$$ \rm L_{IR}$$>$/$\rm <$$\rm L_{8}$$>$, in bins of stellar mass. $\rm <$$\rm L_{IR}$$\rm >$ is estimated by fitting \cite{CE01}, \cite{DH02} and \cite{Rieke2009} templates to stacks of \textit{Herschel}/PACS images, while $\rm <$$\rm L_{8}$$\rm >$ is estimated by fitting these templates to the stacks of 24 \um band images, and \reight is determined as a function of stellar mass. The average bolometric correction factors derived with these three template sets are consistent with each other. The dotted line shows \lir/\leight=4.9 given by \cite{Elbaz2011}.}
\label{fig:lratio_comp}
\end{figure}

\begin{figure*}
\includegraphics[width=0.84\textwidth,center]{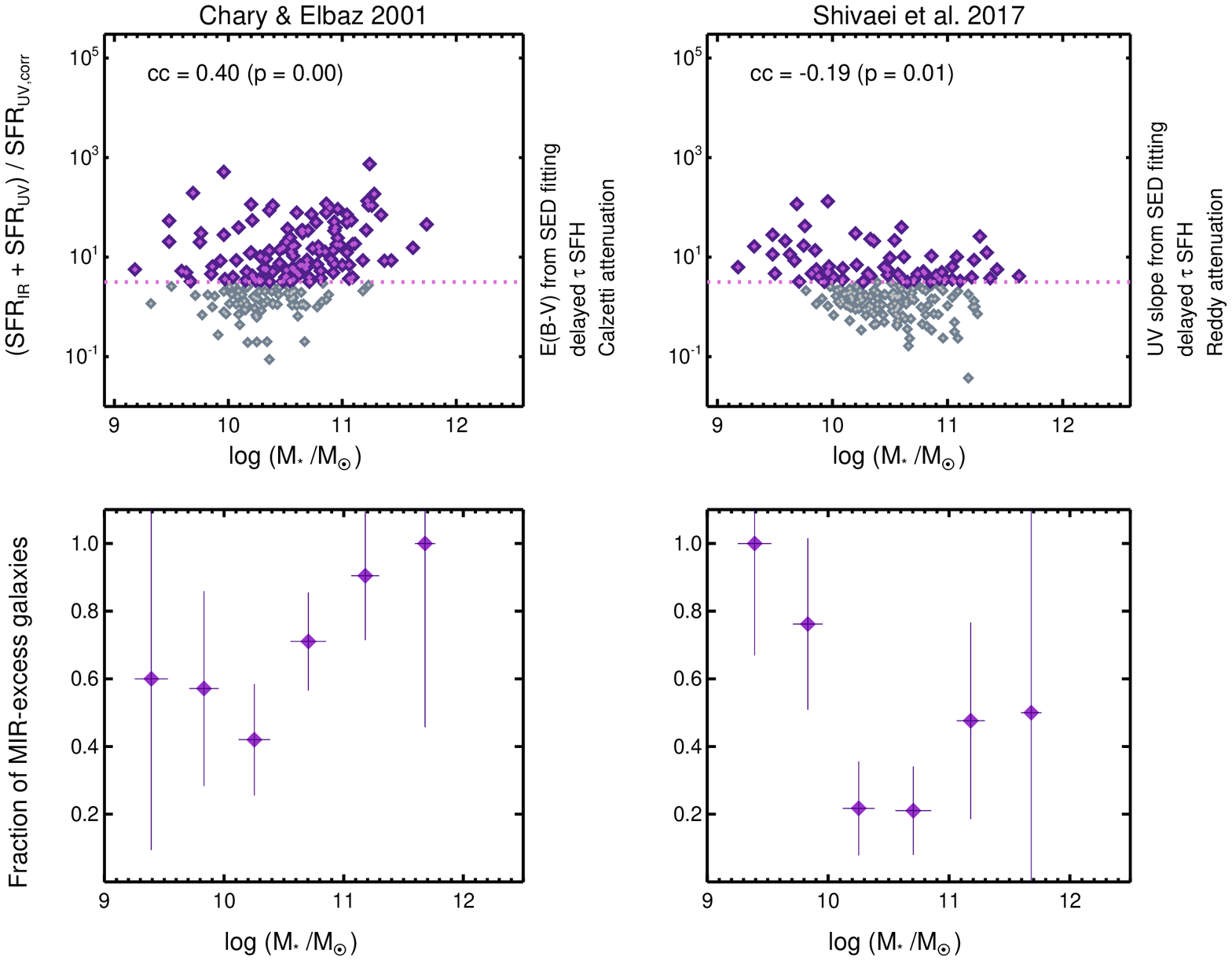}

\caption{\textit{Top:} \sfratio versus stellar mass where \irx are identified using the \db prescription on the left panel and our preferred methodology on the right panel. The correlation coefficients and the significance levels are reported in each panel. \textit{Bottom:} The resulting fraction of \irx identified galaxies in bins of stellar mass, as determined from the top panels. The
different trends observed with stellar mass in these panels is due to a combination of stellar mass selection biases in the survey, inaccurate \lir estimation with \cite{CE01} templates, and variations in the reddening correction factors used.}
\label{fig:mass}
\end{figure*}

In Figure \ref{fig:massdep} we use  $\rm<$$\rm L_{IR}$$>$/$\rm <$$\rm L_{7.7}$$>$ given in bins of stellar mass in \cite{Shivaei2017}, combined with  $\rm L_{7.7}$/\leight=1.25, to estimate \lir for individual galaxies considered in this study.
The median \sfratio ratio indicates an improvement compared to the results shown in the left panel of Figure \ref{fig:lumdep}, which uses the same template set but where \lir was estimated directly from the 24 \um data using the luminosity-dependent from \citet{CE01}.
The percentage of \irx galaxies is substantially lower in Figure \ref{fig:massdep} as well.

Overall, the results using the stellar-mass-dependent method of \citet[see Figure \ref{fig:massdep}]{Shivaei2017} and the luminosity-independent method of \citet[see first column of Figure \ref{fig:lumindep}]{Elbaz2011} are consistent with each other. These methods combined with the \sfuv extinction correction based on the relation between $\rm A_{UV}$ and $\beta$ (estimated from the best fit to the SED assuming a delayed exponentially-declining SFH model, as shown in the third rows of Figures \ref{fig:lumindep} and \ref{fig:massdep}) lead to \sfratio close to unity and identification of 22\%--32\% of our galaxies as MIR-excess. 
We emphasize again that while the similar UV correction method shown in the fourth rows of Figures \ref{fig:lumindep} and \ref{fig:massdep} (which is derived assuming a constant SFH) leads to a much lower fraction of \irx galaxies, a larger dust correction for estimating \sfuvcorr may be required.}


We further test the methodology of \cite{Shivaei2017} with the \cite{DH02} and \cite{Rieke2009} templates (instead of the \cite{CE01} templates) and estimate the average \leight and \lir values in bins of stellar mass. We show \reight in bins of stellar mass in Figure \ref{fig:lratio_comp} and  list the values in Table \ref{tab:irxss3}.
The \reight ratios estimated using these various templates are consistent with each other. While the three sets of templates used for estimating \reight in Table \ref{tab:irxss3} are luminosity-dependent, { we assume that} these ratios are luminosity-independent and can thus be used to estimate \lir directly for a \z galaxy with a 24 \um detection, whether or not it is detected with \textit{Herschel}.
Figure \ref{fig:lratio_comp} also illustrates that the stellar-mass-dependent conversions of \cite{Shivaei2017} are consistent with the single, luminosity-independent conversion factor assumed by \citet{Elbaz2011} for high stellar mass sources ($\log \rm M_*/M_\odot>10$).
\citet{Shivaei2017} find that a larger correction factor is required for lower stellar mass sources, which results in the identification of larger population of \irx galaxies (32\%) with their \sfir estimation compared to the \citeauthor{Elbaz2011} method (22\%). We explore the effect of stellar mass in identification of \irx galaxies further in the next section.
\subsection{MIR-Excess and Stellar Mass} 
\label{sec:stellarmass}

\begin{figure*}
\includegraphics[width=15.1cm,center] {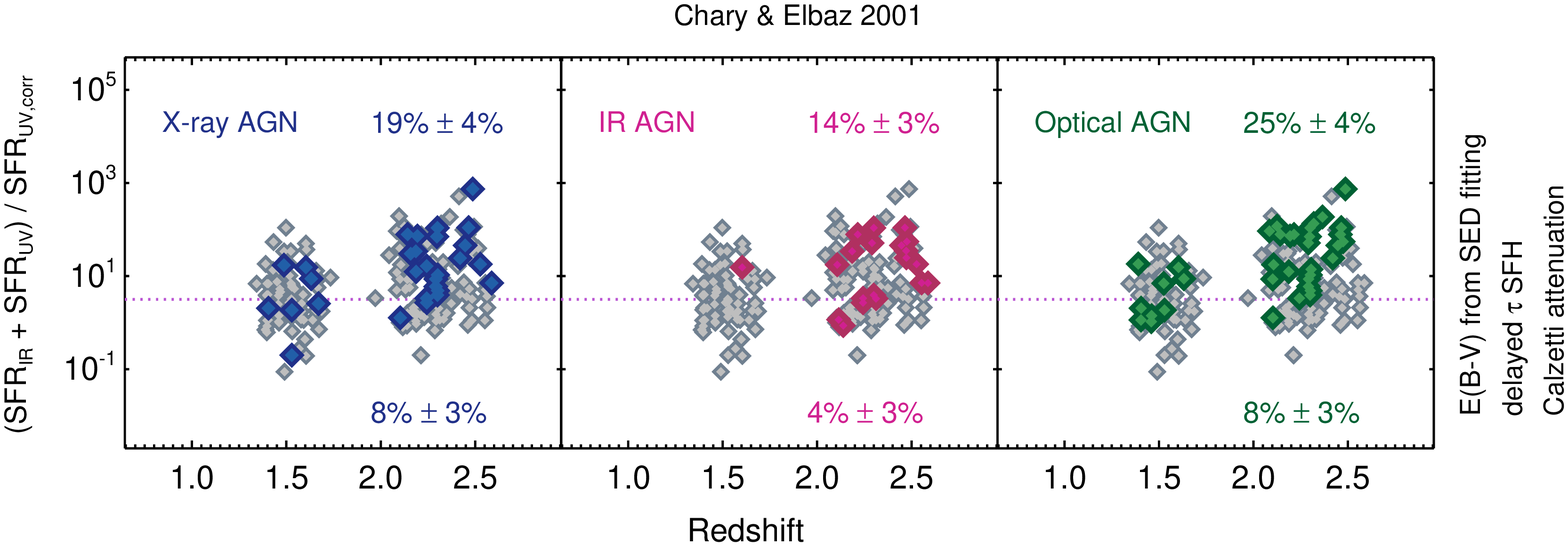}
\\
\\
\includegraphics[width=15.1cm,center] {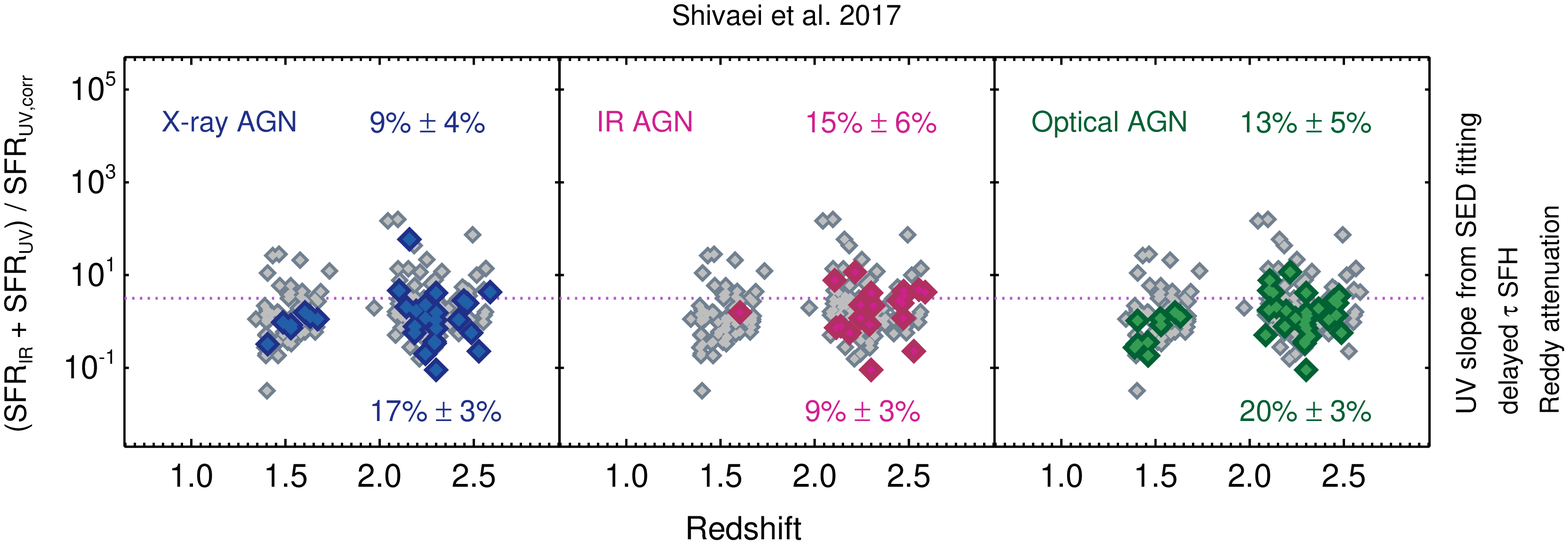}
\caption{
X-ray (left), IR (middle), and optical (right) AGN shown in the \sfratio 
versus redshift space. The fractions in each panel report the percentage of all sources above or below the \irx threshold that are AGN of a given type. \e{In the top panel, for a comparison with \cite{Daddi2007b}, we consider the combination of \cite{CE01} templates with the color-excess estimated from the SED fitting with \cite{calzetti2000} attenuation law. In the top panel, we consider the mass-dependent, luminosity-independent \reight estimates from \cite{Shivaei2017} and \sfuv corrected using the UV spectral slope (our preferred method). With our preferred method, the percentages of AGN above and below the \irx threshold do not reveal a preference for AGN to reside in \irx galaxies, while with a similar prescription to \db there is a preference at 2--3$\sigma$ level which is due to an overestimation of $\rm L_{\rm IR}$.}}
\label{fig:agn}
\end{figure*}

In this section, we examine how the incidence of \irx varies with the stellar mass of the galaxies. In the upper panels of Figure \ref{fig:mass} we plot \sfratio versus stellar mass for \e{the \db prescription (on the left) and for our preferred method (on the right)}, where the stellar masses are derived from SED fitting as discussed in Section \ref{sec:mass}. The correlation coefficients (cc) and their corresponding significance ($p$) are shown in each panel. We use the $r-correlate$ routine in IDL, which calculates the Spearman's rank correlation coefficient and the significance of its deviation from zero. In the bottom panels we plot fraction of \irx galaxies as a function of stellar mass in intervals of $\Delta \log(\frac{{{M_{*}}}}{{{M}_\odot}})$ = 0.5 dex. We calculate the errors assuming a binomial distribution using the Bayesian method of \cite{cameron2011estimation}. 

The relation between \sfratio and stellar mass varies from a statistically significant negative trend in the right panel to a statistically significant positive trend in the left panel. It is possible that neither of these trends reflect an intrinsic, underlying correlation and instead they may be the result of observational selection biases and inaccurate \lir estimates. Figure \ref{fig:mass} shows that there is a selection bias in the lowest mass bin, where we are unable to detect galaxies with low \sfratio. This selection bias combined with an overestimation of \lir in more massive galaxies with \cite{CE01} templates leads to an observed positive correlation in the left panel. In the right panel, for massive galaxies we have a more reliable estimate of \lir and therefore find an overall negative trend due to the selection bias at the low mass end.
If we limit our analysis to the four middle stellar mass bins where the majority of our sources are, we still find a significant difference between the fraction of \irxg in the two panels.

Investigating the relation between \sfir and \sfuvcorr with stellar mass in these two methods can help resolve this discrepancy. The method used by \db overestimates \lir in moderately massive to massive galaxies compared to our preferred method. There is also a large scatter between \sfuvcorr and stellar mass using the \db prescription with the \cite{calzetti2000} attenuation curve, compared to our preferred methodology with the \cite{reddy2015} attenuation curve. A combination of these two trends results in the observed differences in the middle stellar mass bins in Figure \ref{fig:massdep}.

We note that \db find that \irx sources are preferentially identified in galaxies with large stellar masses and find that the fraction of \irxg increases with increasing stellar mass. While using the same methodology we also find a significant positive correlation, \e{and as argued above this could occur due to  selection biases and \lir overestimation.}

Overall, we find that the relation between \sfratio and stellar mass, and the fraction of \irxg in bins of stellar mass, strongly depends on the templates and reddening corrections used. Observational selection biases, inaccurate \lir estimation, and variations in the reddening correction used can lead to the discrepancies between the trends shown in Figure \ref{fig:mass}. 

\subsection{AGN Identified in MIR-Excess Galaxies}\label{sec:detectedagn}

Several studies have proposed that the observed \irx occurs in galaxies due to an underlying AGN contribution \citep[e.g.][]{Daddi2007b,Alexander2011}. \da after excluding the X-ray detected AGN, find that 20\%-30\% of the \z galaxies \e{(BzK selected star-forming galaxies with K $<$20.5)} are \irx  and argue that the existence of this population reflects the presence of an underlying obscured AGN population. Generally, AGN can provide a substantial contribution to the observed MIR emission, though the MIR colors \e{(i.e. IRAC colors)} are distinguishable from star-forming galaxies \citep [e.g][]{alonso2006,donley2012}. Here we address the question of whether AGN may substantially contribute to and perhaps even be the main source of the \irx phenomenon.

\e{While the analysis by \db was limited to X-ray selected AGN}, in the MOSDEF survey we benefit from having a sample of AGN identified at different wavelengths. As discussed in Section \ref{sec:agn}, in addition to X-ray and IR-detected AGN we can use optical BPT selection, which can recover AGN that might be missed by X-ray and IR imaging \citep{Azadi2017}.

\begin{figure*}[t]
\includegraphics[width=0.82\textwidth,center]{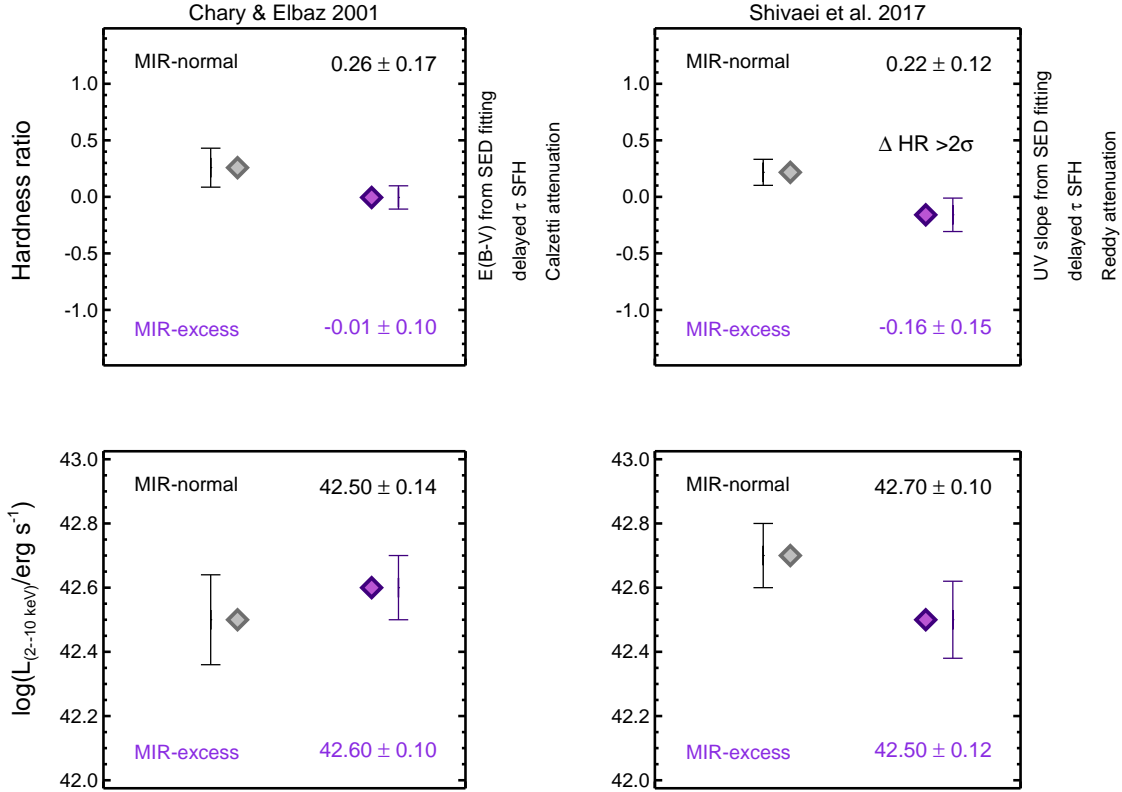}
\caption{\textit{Top:} The hardness ratio of the \irx (purple) and MIR-normal galaxies (gray), after excluding X-ray detected AGN, \e{using the same prescription as \db on the left side and our preferred method on the right side.} We do not find a significantly harder X-ray spectrum in \irx galaxies compared to MIR-normal galaxies in any of these panels, which argues that AGN are not substantially contributing to the observed  MIR-excess. \textit{Bottom:} The average X-ray luminosities for the MIR-normal and \irx galaxies. The X-ray luminosities are consistent with star formation origin in both populations, in the both panels.} 
\label{fig:xstack}
\end{figure*}

In Figures \ref{fig:lumdep} to \ref{fig:massdep} we presented 24 different methods for the identification of \irx galaxies, based on different combinations of methods to correct the UV emission for dust and extrapolate from the observed 24 \um flux to estimate the total IR luminosity. As discussed above, the combination of the dust correction estimated from the UV spectral slope and the \cite{Shivaei2017} method (the luminosity-independent, mass-dependent, \reight derived using \textit{Herschel} stacks)  \e{results in  $\rm SFR_{IR}+SFR_{UV}$ and $\rm SFR_{UV,corr}$ being consistent for the bulk of the galaxies} and is therefore the preferred methodology in this paper. \e{
The prescription used by \db corresponds to using the color-excess from the SED fits with the \citeauthor{calzetti2000} reddening curve and the \cite{CE01} luminosity-dependent IR templates. We compare our preferred methodology and the \db prescription in Figure \ref{fig:agn} and find the fraction of sources that are identified as X-ray, IR, or optical AGN above and below the \irx threshold.} \e{With the \db prescription in the top panel of Figure \ref{fig:agn} we find a higher percentage of X-ray, IR, and/or optical AGN at $2-3\sigma$ in the \irx region.}  With our preferred method in the bottom panel, we find that none of the three AGN identification techniques result in AGN being more prevalent in the \irx region. However, the median \sfratio ratio of 4.9 found with the \db prescription  indicates that the higher incidence of AGN in \irx galaxies is due to the faulty estimation of the total SFR, and with the more reliable estimates in the bottom panel of this figure we no longer find any evidence for a significantly higher prevalence of AGN in \irx galaxies.

We note that while in Figure \ref{fig:agn} we  illustrate only two of the 24 methods of SFR estimation discussed in this paper, we perform a similar analysis for all the other methods as well. 
{ Of the other 22 combinations of methods investigated for SFR estimation,}
none results in a significantly higher fraction (at even the 2$\sigma$ level) of AGN among \irx galaxies than in MIR-normal galaxies. 


\subsection{X-ray Stacking Analysis of MIR-Excess Galaxies}
\label{sec:stacks}

X-ray detection is a well-known, reliable method for robust AGN identification \citep[e.g.][]{mendez2013primus,Azadi2017}, however it fails to identify highly absorbed AGN with hydrogen column densities of $N_{H}> 1.5 \times 10^{24}$ cm$^{-2}$ \citep[e.g.][]{DellaCeca2008,comastri2011}. Studies predict that 10-50\% of AGN are Compton thick \citep[e.g.][]{Akylas2009,Alexander2011,Lanzuisi2015}. The absorbed and reprocessed radiation from these AGN buried in dust may contribute to the observed MIR radiation in galaxies.

To examine whether the excess observed at MIR wavelengths in our galaxies occurs due to obscured AGN activity, we exclude the X-ray detected AGN from our sample and then stack the X-ray photons from the remaining sources. 
In this section we investigate the hardness ratio (H-S/H+S), where H and S are  the net count rates in the hard (rest-frame 2-10 keV) and soft (rest-frame 0.5-2 keV) X-ray bands, respectively, and are calculated for both the \irx and MIR-normal galaxies. The hardness ratio is a tracer of the X-ray spectrum shape and the amount of obscuration. A positive hardness ratio indicates sources with a hard X-ray spectrum, indicating obscured  AGN, while a negative hardness ratio indicates sources with a softer spectrum, unobscured AGN or the X-ray emission from star-forming galaxies. To estimate H and S, we use the number of counts measured within an aperture with a radius corresponding to 70\% of the enclosed energy fraction for the PSF at each galaxy position, while the background is determined from background maps based on smoothing of local counts. 

In Figure \ref{fig:xstack} we present the hardness ratio for the \irx (purple) and MIR-normal (gray) galaxies \rf{using the \db } method (left) and our preferred method (right). The errors on the hardness ratios are calculated using bootstrap resampling. We find a relatively harder signal in MIR-{\it normal} galaxies compared to the \irx population (at the 2$\sigma$ level with our preferred method). The negative hardness ratios indicate that a soft X-ray signal dominates in \irx galaxies. When we consider all the other combinations of methods as discussed above, in none of them does the hardness ratio of \irxg reflect  
an underlying hard spectrum which could be due to an AGN. Finding a harder spectrum in MIR-normal galaxies compared to \irx galaxies is at odds with the findings of \db.

We also calculate the average X-ray luminosities for the MIR-normal and \irx galaxies based on our X-ray stacking in the hard X-ray band. When using our preferred method to identify \irx galaxies, \e{we measure an average rest-frame 2--10~keV X-ray luminosity of $\log (\rm L_{2-10keV}/{erg\;s^{-1}})={42.7\pm0.10}$ for the MIR-normal galaxies.} This luminosity is higher than the expected X-ray emission due to star formation processes: $\log (\rm L_{SF}/\mathrm{erg\;s^{-1}})\approx42$ for $z\sim2.1$ star-forming galaxies with stellar masses $\sim 10^{10.5} M_\odot$, typical for our sample
\citep[see][]{Lehmer2016,Aird2017}. 

\e{For the MIR-excess galaxies we find a slightly \emph{lower} average luminosity, $\rm \log (L_{2-10keV}/{erg\;s^{-1}})={42.5\pm0.12}$. We note that while the luminosities of both of these populations are consistent with a star formation origin, the slightly higher $\rm L_{(2-10keV)}$ in both \irx and MIR-normal galaxies compared to the main sequence of star formation (for a similar stellar mass and redshift regime ) could be due to some contribution from X-ray undetected AGN in this sample, as might be expected. The slightly lower $\rm L_{X}$ in \irx galaxies compared to the MIR-normal galaxies may be due  to the slightly lower masses of \irx galaxies ($\rm 1.7\times10^{10} M_{\sun})$ compared to the MIR-normal galaxies ($ \rm  2.9\times10^{10} M_{\sun}$). } 

\e{Using the \db prescription we find $\rm \log (L_{2-10keV}/\mathrm{erg\;s^{-1}})={42.5\pm0.14}$ for the MIR-normal galaxies and  $\rm \log (L_{2-10keV}/\mathrm{erg\;s^{-1}})={42.6\pm0.10}$ for \irx galaxies. The  luminosities and the hardness ratios of the MIR-normal and \irx galaxies identified with the \db method are both consistent with a star formation origin.} The hard spectrum of the \irx galaxy sample originally identified by \db is probably due to a few undetected X-ray AGN, which would be detected with deeper X-ray surveys \citep[see also][]{Alexander2011,Rangel2013}.

We showed in the section above that there is not a higher fraction of detected AGN above the \irx threshold than below it with our preferred method. Here we find that a stacking analysis of our X-ray undetected sample does not indicate a harder spectrum in \irx galaxies than in MIR-normal galaxies, which shows that AGN do not contribute substantially to the MIR radiation in our \irx sample.

\begin{figure*}
\includegraphics[width=0.89\textwidth,center]{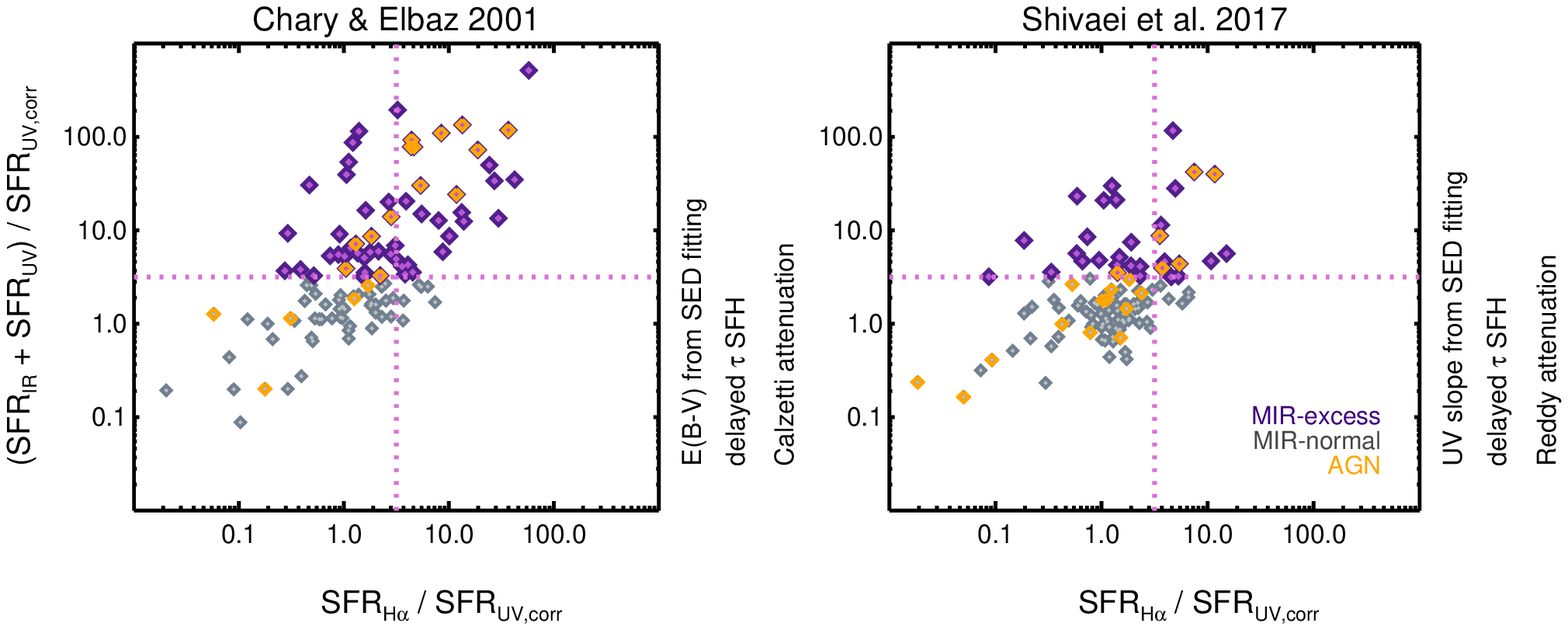}

\caption{\e{\sfratio versus \sfha/\sfuvcorr for the galaxies and AGN with significant detection at 24 \um band. The horizontal dotted line indicates the \sfratio $\sim 3.2$ defined by \db for \irx galaxies identification, while the vertical dotted line indicates \sfha/\sfuvcorr $\sim 3.2$. AGN are included and shown in orange. \irx galaxies identified either with the \db methodology or our preferred methodology do not necessarily have \sfha/\sfuvcorr beyond the vertical limit.}}

\label{fig:ha}
\end{figure*}



\subsection{ $SFR_{H\alpha}$ of the MIR-Excess Galaxies} \label{sec:sfha}

In MOSDEF we benefit from having spectroscopic measurements for a statistically large sample of galaxies at $z\sim2$. 
Therefore, we can also estimate the SFR based on the \ha\ emission line (hereafter, \sfha)
and investigate whether \irx galaxies have elevated \sfha compared to the MIR-normal galaxies.

\e{In Figure \ref{fig:ha} we illustrate \sfratio versus \sfha/\sfuvcorr for the galaxies and AGN in our sample. For this analysis we limit our sample to the sources with significant ($>3\sigma$) \ha\; and \hb\; detections, which decreases our sample size to 114 sources. \sfha is calculated from the luminosity of the \ha\ line using the relation given by \cite{kennicutt1998star} and is corrected for  attenuation using the Balmer decrement (\ha/\hb). Similar to the plots shown above, AGN are included in this analysis and are shown in orange, although the AGN may contribute to the \ha\ emission. The horizontal dotted line is the threshold of \sfratio $\sim 3.2 $ defined by \db for \irx identification.
The vertical dotted line shows the equivalent threshold on \sfha/\sfuvcorr $\sim3.2$, identifying sources that exhibit an excess in \ha\ emission (relative to the dust-corrected UV continuum emission).}

\e{In both the \db prescription (left) and our preferred methodology (right) we find a positive trend between \sfratio and \sfha/\sfuvcorr. In both panels the majority of the MIR-normal galaxies also have \sfha/\sfuvcorr less than the vertical threshold, and only a very few of the MIR-normal galaxies are beyond this limit. The majority of \irx galaxies identified with our prescription are below the vertical limit in the right panel, while with the \db prescription in the left panel almost half of them are above the vertical limit. However, as discussed above such a high \sfratio  using \db prescription is due to an overestimation of the total IR luminosity. We also note that almost 40\% of the \irx sources that are beyond the vertical limit are AGN in both panels and are expected to have elevated \ha\ emission.}

\e{Overall, Figure \ref{fig:ha} shows that \irx galaxies in MOSDEF do not necessarily have high \sfha/\sfuvcorr values.}\e{ Therefore, while the elevated \sfir in $\sim30\%$ of the galaxies in our sample may 
be due to legitimately higher \sfir than other SFR tracers, it may also be the result of an overestimation of \sfir.}

\section{Discussion}
\label{sec:discussion}

In this paper we use data from the MOSDEF survey to investigate the nature of \irx galaxies at $z\sim2$.  We find that the \irx incidence depends strongly on the templates used for estimating \lir and the reddening correction of \sfuv, with the percentage of galaxies identified as \irx varying between $\sim$10-80\% depending on which of these methods is used. 
\e{Using the mass-dependent, luminosity independent templates from \cite{Shivaei2017} for estimating \lir and the UV slope for the \sfuv correction, we identify  32\%$\pm$3\% of} the MOSDEF galaxies as \irx galaxies. \e{We note that while this combination of \sfir and \sfuv results in \sfratio close to unity, there may be some galaxies for which this ratio is legitimately greater than one, though we do not expect this for the bulk of the population}.

In this section we first discuss how the specific method used to determine the UV reddening correction plays an important role in robustly  estimating the extinction-corrected \sfuv.  We then discuss how PAH features may contribute to the observed MIR-excess. We then investigate the AGN contribution to \irx galaxies, and finally, we compare the identification of \irxg in this study with \db.


\subsection{UV Reddening Correction Estimation} \label{discuss1}

\e{In Section \ref{sec:results} (Figures \ref{fig:lumdep} -- \ref{fig:massdep}) we examine various  methods for the reddening correction of the UV photometry: using the color-excess estimated directly from  SED fitting with the \cite{calzetti2000} and \cite{reddy2015} attenuation curves, and using the average relationship between reddening and UV continuum slope 
assuming either a delayed-$\tau$ or constant SFH model for the underlying stellar population.
Our results show that the different methods for estimating the reddening correction result in substantial changes to the fraction of \irxg identified (for a given set of IR templates).}

\e{Using the attenuation curve of \cite{reddy2015} results in a decrease in \sfratio and the fraction of \irx, compared to using the widely adopted \cite{calzetti2000} attenuation curve. 
The magnitude of the decrease depends on the \lir templates used. 
As discussed in \cite{reddy2015} the overall shape of the \citeauthor{calzetti2000} and \citeauthor{reddy2015} attenuation curves are identical at short wavelengths, but the \citeauthor{reddy2015} curve has a lower normalization. The substantially redder E(B-V) ($\Delta$E(B-V)$\sim0.1$) of 
\citeauthor{reddy2015}\footnote{We note that $\Delta$E(B-V) of the \cite{reddy2015} attenuation curve compared to that of \cite{calzetti2000} may also depend on the adopted SFH.} results in a higher UV correction factor compared to \citeauthor{calzetti2000} and hence a lower fraction of \irx galaxies.}

We find that using the UV spectral slope for the reddening correction, along with robust estimates of \lir, results in a median \sfratio ratio close to unity, illustrating a consistency between the $\rm SFR_{IR}+SFR_{UV}$ and $\rm SFR_{UV,corr}$ measurements. \e{Using the UV slope also results in the lowest percentage of \irxg ($12-74\%$ depending on the \lir estimation and the adopted SFH).}

\begin{figure}\includegraphics[width=0.40\textwidth,center] {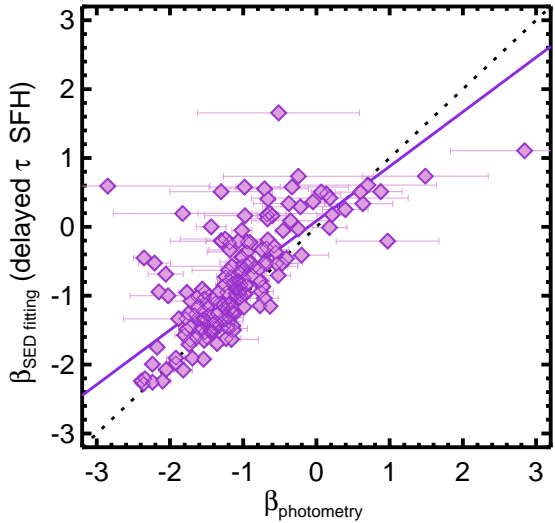}
\caption{The UV spectral slope estimated from fitting a power-law to the full SED fit ($\rm \beta_{SED}$) versus the slope from fitting a power-law to the photometric data at 1200-2600 \AA\; ($\rm \beta_{phot}$). The gray dotted line is the 1:1 line, and the gray solid line is the best fit line ($\rm \beta_{SED}=0.8\beta_{phot}-0.1$). The large uncertainties on the photometric data lead to the large error bars on ($\rm \beta_{phot}$). On average $\rm \beta_{SED}$ is larger than $\beta_{phot}$, which results in a higher reddening correction to \sfuv and consequently a lower \sfratio and lower fraction of \irxg compared to results using $\rm \beta_{phot}$.}
\label{fig:beta}
\end{figure}

The UV slope used in this analysis is measured by fitting a power law to the SED fit ($\rm \beta_{SED}$) at 1200-2600 \AA\;  \e{assuming a delayed-$\tau$ SFH model}(see Section \ref{uvcorr}). However, it is also common to estimate the UV slope by fitting a power law directly to the photometric data ($\rm \beta_{phot}$) at these wavelengths \citep[e.g.][]{reddy2015,Shivaei2015}. 
\e{We find for the MOSDEF sample that using $\rm \beta_{phot}$ instead of $\rm \beta_{SED}$ increases the fraction of \irxg from $32\%\pm3\%$ to $50\%\pm3\%$ and the median \sfratio to 3.2 \citep[using][\lir estimates]{Shivaei2017}.} In Figure \ref{fig:beta} we plot $\rm \beta_{phot}$ versus $\rm \beta_{SED}$ and find that while the values roughly scatter around the 1:1 line, the best fit line is $\rm \beta_{SED}=0.8 \beta_{phot}+0.1$.  This results in a larger extinction correction factor ($10^{0.4\rm A_{UV}}$, \e{assuming Equation \ref{eq:auvbeta}}) using $\rm \beta_{SED}$, and consequently a smaller fraction of \irx galaxies are identified using $\rm \beta_{SED}$ instead of $\rm \beta_{phot}$. 

\e{The error bars in Figure \ref{fig:beta} demonstrate the large uncertainty in the $\rm \beta_{phot}$ measurements, due to the relatively large errors on the UV photometric data, particularly in red galaxies that are faint at UV and optical wavelengths. Despite the large errors, $\rm \beta_{phot}$ provides an estimation of the slope independent of any assumption about the stellar population, while $\rm \beta_{SED}$ relies on the models adopted in the SED fitting. However, this can be a disadvantage in  galaxies that have redder slopes due to an older stellar population rather than dust obscuration. We note that while the errors on  $\rm \beta_{SED}$ are not calculated here they are expected to be smaller than the errors on $\rm \beta_{phot}$ as they are estimated using many more photometric data points.}

Overall, we find that $\rm \beta_{SED}$ provides \e{a good estimate of \sfuvcorr when comparing \sfuvcorr with \sfuv$+$\sfir, and the commonly-used $\rm \beta_{phot}$ may underestimate the reddening correction factor as it results in substantially higher fraction of \irx galaxies and average \sfratio}. Fitting a power law to the SED to estimate the UV spectral slope may be more reliable than fitting to the photometry directly, due to the larger uncertainties in the UV photometric data. 


\subsection{$L_{IR}$ Estimation and PAH Contribution}
\label{sec:discuss2}

In this study we considered the use of six different templates/correction factors to estimate the total \lir for the MOSDEF galaxies from the observed 24~\um flux. We find that extrapolating from the observed 24 \um data using templates based on local galaxies results in an overestimation of \lir \citep[see also][]{Reddy2012} and consequently overestimates the fraction of \irx galaxies. Our results using the methodologies of  \cite{Elbaz2011} and \cite{Shivaei2017}, 
considering luminosity-independent corrections
which were calibrated using stacks of the FIR data,
leads to a more accurate estimation of \lir \e{\citep[as also seen in ][among others]{Santini2009,Elbaz2010,Nordon2010,Rodighiero2010,Elbaz2011}}. Using the methods of  \cite{Elbaz2011} and \cite{Shivaei2017} (along with the UV spectral slope)  we identify 22\%$\pm$3\% and  32\%$\pm$3\% of our galaxies, respectively, as \irx galaxies; these fractions are lower than those obtained using other IR templates. The advantage of \cite{Shivaei2017} is that it takes into account the dependence  of \reight on stellar mass. 
As illustrated in Figure \ref{fig:lratio_comp} the ratios using the \cite{Shivaei2017} method (with the template sets of \cite{CE01}, \cite{DH02} or \cite{Rieke2009}) are generally consistent with \lir/\leight from \cite{Elbaz2011} and are discrepant only in our lowest stellar mass bin of 
$\rm M_{*} < 10^{10} M_{\sun}$.

\e{Using a robust \lir estimation and  the UV slope for the reddening correction, we find that $\sim30\%$ of the galaxies in our sample are MIR-excess. Our analysis in Sections \ref{sec:detectedagn} and \ref{sec:stacks} does not provide any evidence for substantial AGN contribution in \irx galaxies compared to the MIR-normal galaxies. Therefore, the excess MIR radiation  detected in $\sim$30\% of our sources is likely due to other phenomena contributing to MIR radiation. 
The observed 24 \um band at \z  also includes  line emission from PAH molecules at rest-frame wavelengths of 5--12 \um. Therefore, the incidence of \irx may be  due to PAH features in normal star-forming galaxies entering the infrared bands \citep[see also][]{Nordon2010,lutz2011pacs,Nordon2012}}

While studies predict 5--20\% of the bolometric IR radiation is from PAH molecules \citep[e.g.][]{Smith2007,Shipley2016}, the strength of radiation from PAH molecules varies with galaxy stellar mass, metallicity, and/or star formation activity \citep[e.g.][]{Engelbracht2006,Smith2007,Shipley2016, Shivaei2017}.
\ma{\cite{Shivaei2017}, after including PAH features in their templates and determining calibrations from \textit{Herschel} data, find that the PAH radiation at 7.7 \um depends on the age and stellar mass of galaxies and is lower in young galaxies with ages $\lesssim$ 500 Myr as well as galaxies with $\rm M_{*} < 10^{10} M_{\sun}$.} \ma{Our analysis in the right panel of Figure \ref{fig:mass} indicates  that at $\rm M_{*} \lesssim 10^{10} M_{\sun}$ we  identify many \irx galaxies. This high incidence of \irx in low mass galaxies could be due to a selection effect: given that our sample is selected based on 24 \um observations, if a galaxy has strong PAH features it is more likely to be detected at 24 \um and to be included in our sample. While according to \cite{Shivaei2017} in lower mass galaxies, for a given \lir, PAH emission is suppressed, there could also be an increased scatter in the strength of PAH emission in individual galaxies at lower mass.  Such an increased scatter could result in a higher fraction of \irx galaxies at $\rm M_{*} \lesssim 10^{10} M_{\sun}$.}

\ma{Indeed, the scatter in the relative strength of PAH emission to \lir is lower in moderate mass galaxies with $\rm M_{*} > 10^{10} M_{\sun}$  \citep{Shivaei2017}.} If we limit our analysis to galaxies with $\rm M_{*} > 10^{10} M_{\sun}$, where we likely do not have stellar mass selection biases in our sample, the fraction of \irx galaxies decreases to $24\% \pm 2\%$. However, given that \sfha is not elevated in the \irx galaxies in our sample, the intrinsic fraction of \irx may be even lower than 24\%.


\subsection{AGN Contribution}

Our multi-wavelength AGN and X-ray stacking analysis shows that it is unlikely for either detected or undetected AGN to contribute substantially to the MIR excess observed in our \irx galaxies. 
\e{Following a similar prescription as \db for SFR estimations, we find 
that X-ray AGN are more prevalent in the \irx region at the $2\sigma$ level and that IR and optical AGN show the same behavior as well (despite the fact that unlike \db our sample is not BzK-selected).}  However, when using more robust estimates of SFRs we do not find a higher incidence of AGN in the \irx region; therefore the higher prevalence of detected AGN in \irx sources using the methodology of \db primarily occurs due to inaccurate SFR estimates.

More recently, using similar methodologies as \db, 
 \rang find that at \z the fraction of X-ray detected AGN above the \irx threshold is three times higher than below the threshold. However, they argue that this is due to a strong correlation between \sfratio and \lir and show that the fraction of X-ray detected AGN above and below the \irx threshold is very similar if considered in narrow bins of \lir.

Our stacking analysis in Figure \ref{fig:xstack} shows that the hardness ratio of X-ray undetected \irx and MIR-normal galaxies \e{does not vary depending on the methods used for to estimate SFRs}. Unlike in \db, in almost all of our combinations of methods we find a harder X-ray spectrum for MIR-normal galaxies than for the \irx population,
\e{indicating, if anything, a higher incidence of AGN in the \emph{normal} galaxies than the \irx population.}
With our preferred methodology  (right column of Figure \ref{fig:xstack}) we find that the \irx galaxies have a  soft X-ray spectrum with a hardness ratio of $-0.16\pm0.14$ and an average X-ray luminosity of $\sim10^{42}$ erg/s in both the soft and hard X-ray bands,
\e{consistent with a minimal AGN contribution and the bulk of the X-ray emission from the \irx galaxies being associated with star formation.}

Using deep X-ray data in the CDFN and CDFS fields, \rang stack  the X-ray undetected sources and find hardness ratios of $-0.50\pm0.12$ and  $-0.45\pm0.15$ respectively above and below the \irx threshold. Neither of these hardness ratios indicate the presence of underlying obscured AGN activity. By limiting their sample to the same depth (1 Ms) as \db, \rang find a  harder spectrum with a hardness ratio of  $-0.16\pm0.16$ for \irx galaxies and argue that the relatively harder spectrum in \db is due to a few undetected X-ray AGN in that sample that are detected with the deeper X-ray data. These findings indicates that some of the sources identified as \irx  in the \db sample are indeed obscured AGN, but this does not show that an \irx occurs generally due to a contribution from AGN.

While the \db and \rang samples utilize the CDFS and CDFN fields  \citep[see also][]{Alexander2011}, MOSDEF spans a larger number of fields, which have different X-ray depths. To investigate the effect of X-ray depth we estimate the hardness ratio of the X-ray undetected sample in each of our fields above and below the \irx threshold, using our preferred method for SFR estimates. We do not find a harder spectrum in \irx stacks compared to the MIR-normal galaxies in any of these fields. Overall, our stacking analysis does not indicate a hard X-ray spectrum or a strong AGN contribution to \irx galaxies. 

\subsection{Comparison with \db Results} 

\e{In this study using our preferred methodology we identify $32\%\pm3\%$ of the MOSDEF sample as \irx galaxies. As discussed above in Section \ref{sec:discuss2}, once the stellar mass selection biases are taken into account, this fraction decreases to $24\%\pm2\%$. While these fractions are similar to the fraction of \irx galaxies ($\sim$20-30\%) identified by \db in their BzK-selected sample at a similar redshift and a similar stellar mass regime, as discussed extensively above the overestimation of \lir in the \db analysis is likely to play an important role in their identification of \irxg.}Using a similar approach to \db (in column 1-row 1 of Figure \ref{fig:lumdep}) we identify $61\%\pm3\%$ of our sources as \irx galaxies. 

\e{We note  additional minor differences in our approaches: 
here we use the color-excess estimated from SED fitting with the \cite{calzetti2000} attenuation curve for the reddening correction in our estimate of \sfuvcorr, whereas \db use the relation between E(B-V) and B-z color from \cite{Daddi2004} assuming the \cite{calzetti2000} attenuation curve. In addition, the \db sample is BzK-selected and different galaxy selections could lead to differences in the \irx fraction.
With a similar \lir estimation as \citeauthor{Daddi2007b}, using our UV spectral slope method instead of the \citeauthor{calzetti2000} attenuation curve results in a significant decrease in the \irx fraction in Figure \ref{fig:lumdep}. 
Therefore, the identification of a large population of \irx galaxies when reproducing the \db method appears to be due to underestimating the dust-corrected \sfuvcorr combined with an overestimate of \sfir due to the use of locally-calibrated, luminosity-dependent IR templates.}

\e{With a more careful estimation of the SFRs, and after correcting for stellar mass selection biases, in this work we find that 24\% of our \z  galaxies are MIR-excess.  This incidence is primarily due to  enhanced PAH emission at $z\sim2$ in moderate mass galaxies, rather than obscured AGN activity. PAH emission may also affect the incidence of \irx in the \citeauthor{Daddi2007b} sample but the inaccurate SFR estimation is likely to play a dominant role in their analysis.}

\section{Summary}
\label{sec:summary}

In this paper we use the data from the first three years of the MOSDEF survey to study the nature of \irx galaxies. We define \irx galaxies as those where the combined \sfir and \sfuv exceeds the extinction-corrected \sfuv by a factor of three or greater. We limit our analysis to  galaxies with significant 24 \um and \e{UV} detections, and to ensure that the 24 \um data trace the rest-frame 8 \um data we further restrict our analysis to the 194 galaxies and AGN at $1.40<z<2.61$. We investigate  templates from various studies for \lir estimation as well as various reddening correction methods, and identify the preferred combination as the one where the SFR estimates agree such that \sfir + \sfuv $\sim$ \sfuvcorr. We investigate how different combinations of templates and reddening corrections result in identification of a different fraction of our sample as \irx galaxies. We further investigate the contribution from AGN in the excess MIR radiation. 

Our main conclusions are as follows:

\begin{itemize}

\item The identification of \irx galaxies is strongly dependent on the UV dust reddening correction method used as well as the total IR luminosity estimation. Our preferred methodology \citep[\sfir estimation using results from][and UV slope estimation from the best-fit SED assuming a delayed-$\tau$ SFH for \sfuvcorr estimation]{Shivaei2017}  results in identification of $\sim30\%$ of the MOSDEF sample as \irx galaxies. \e{This fraction decreases to 24\% once  stellar mass selection biases are taken into account.}

\item The bolometric IR luminosity estimated from locally-calibrated, luminosity-dependent templates overestimates \lir and \sfir in \z galaxies. Our analysis shows that the stellar mass-dependent, luminosity-independent method of \cite{Shivaei2017}, which is calibrated \ma{including PAH emission and} using stacks of \textit{Herschel} FIR data, provides a more robust \lir estimation at higher redshifts and results in the identification of fewer galaxies as MIR-excess sources.

\item The UV spectral slope estimated by fitting a power-law to the SED fit at UV wavelengths provides a reliable reddening correction in that it results in consistency between  \sfuvcorr and \sfir +  $\rm SFR_{UV}$. The UV spectral slope can also be measured by fitting a power-law directly to the photometric data, though this has larger errors and leads to the identification of a higher fraction of galaxies as MIR-excess galaxies. 

\item The \sfha in \irx galaxies is not elevated compared to MIR-normal galaxies in our sample, such that an overestimation of \sfir may result in identification of some  galaxies as MIR-excess in our sample. \e{Therefore, the intrinsic fraction of \irx may be even lower than 24\%.}

\item Using reliable estimates of \sfir and dust-corrected $\rm SFR_{UV}$, we do not find a higher fraction of AGN detected in \irx galaxies compared to MIR-normal galaxies. Additionally, stacking the X-ray undetected galaxies does not reveal a hard X-ray spectrum in \irx galaxies. Therefore AGN are not the dominant cause of galaxies having an MIR-excess.

\item The \irx phenomenon \e{in our moderate mass galaxies ($ \rm M_{*} > 10^{10} M_{\sun}$)} is  most likely due to the enhanced emission from PAH dust molecules as 24 $\um$ band at \z traces  PAH features.

\end{itemize}

\acknowledgements
We thank the MOSFIRE instrument team for building this powerful instrument. Funding for the MOSDEF survey is provided by NSF AAG grants AST-1312780, 1312547, 1312764, and 1313171 and grant AR-13907 from the Space Telescope Science Institute. We acknowledge Mark Dickinson and Hanae Inami for providing part of the IR data used in this work.
JA acknowledges support from ERC Advanced Grant FEEDBACK 340442. The data presented herein were obtained at the W. M. Keck Observatory, which is operated as a scientific partnership among the California Institute of Technology, the University of California and the National Aeronautics and Space Administration. The Observatory was made possible by the generous financial support of the W. M. Keck Foundation. The authors wish to recognize and acknowledge the very significant cultural role and reverence that the summit of Mauna Kea has always had within the indigenous Hawaiian community. We are most fortunate to have the opportunity to conduct observations from this mountain.

\bibliographystyle{apjurl}
\bibliography{references.bib}

\maketitle

\end{document}